\documentclass[iop,revtex4,numberedappendix]{emulateapj}
\usepackage{graphicx}
\usepackage{natbib}
\usepackage{amssymb}
\usepackage{amsmath}
\usepackage{xcolor}
\usepackage[hyperindex,breaklinks,bookmarks=false]{hyperref}
\usepackage[figuresright]{rotating}

\def\lesssim{\lower.5ex\hbox{$\; \buildrel < \over \sim \;$}}
\def\gtrsim{\lower.5ex\hbox{$\; \buildrel > \over \sim \;$}}

\shorttitle{The Stellar Populations of Ultra-diffuse Galaxies}
\shortauthors{V. Pandya et al.}

\date{\today}

\begin{document}

\title{The Stellar Populations of Two Ultra-diffuse Galaxies from Optical and Near-infrared Photometry}
  
\author {Viraj Pandya$^{1}$, Aaron J. Romanowsky$^{2,3}$, Seppo Laine$^{4}$, Jean P. Brodie$^{1,3}$, Benjamin D. Johnson$^{5}$, William Glaccum$^4$, Alexa Villaume$^1$, Jean-Charles Cuillandre$^6$, Stephen Gwyn$^7$, Jessica Krick$^4$, Ronald Lasker$^8$, Ignacio Mart{\'i}n-Navarro$^{1,3}$, David Martinez-Delgado$^{9}$, Pieter van Dokkum$^{10}$}
\affil{$^{1}$Department of Astronomy and Astrophysics, University of California, Santa Cruz, CA 95064, USA}
\affil{$^{2}$Department of Physics \& Astronomy, San Jos\`{e} State University, One Washington Square, San Jose, CA 95192, USA}
\affil{$^{3}$University of California Observatories, 1156 High Street, Santa Cruz, CA 95064, USA}
\affil{$^{4}$IPAC, Mail Code 314-6, Caltech, 1200 E. California Blvd., Pasadena, CA 91125, USA}
\affil{$^{5}$Harvard-Smithsonian Center for Astrophysics, 60 Garden St., Cambridge, MA 02138, USA}
\affil{$^{6}$CEA/IRFU/SAp, Laboratoire AIM Paris-Saclay, CNRS/INSU, Université Paris Diderot, Observatoire de Paris, PSL Research University, F-91191 Gif-sur-Yvette Cedex, France}
\affil{$^{7}$Herzberg Institute of Astrophysics, National Research Council of Canada, Victoria, BC V9E 2E7, Canada}
\affil{$^{8}$Finnish Centre for Astronomy with ESO (FINCA), University of Turku, V{\"a}is{\"a}l{\"a}ntie 20, FI-21500 Kaarina, Finland}
\affil{$^{9}$Astronomisches Rechen-Institut, Zentrum f{\"u}r Astronomie, Universit{\"a}t Heidelberg, M{\"o}nchhofstr. 12--14, 69120 Heidelberg, Germany}
\affil{$^{10}$Astronomy Department, Yale University, New Haven, CT 06511, USA}
\email{viraj.pandya@ucsc.edu}  

\begin{abstract}
We present observational constraints on the stellar populations of two ultra-diffuse galaxies (UDGs) using optical through near-infrared (NIR) spectral energy distribution (SED) fitting. Our analysis is enabled by new $Spitzer$-IRAC 3.6 $\mu$m and 4.5 $\mu$m imaging, archival optical imaging, and the \texttt{prospector} fully Bayesian SED fitting framework. Our sample contains one field UDG (DGSAT I), one Virgo cluster UDG (VCC 1287), and one Virgo cluster dwarf elliptical for comparison (VCC 1122). We find that the optical--NIR colors of the three galaxies are significantly different from each other. We infer that VCC 1287 has an old ($\gtrsim7.7$ Gyr) and surprisingly metal-poor ($[Z/Z_{\odot}]\lesssim-1.0$) stellar population, even after marginalizing over uncertainties on diffuse interstellar dust. In contrast, the field UDG DGSAT I shows evidence of being younger than the Virgo UDG, with an extended star formation history and an age posterior extending down to $\sim3$ Gyr. The stellar metallicity of DGSAT I is sub-solar but higher than that of the Virgo UDG, with $[Z/Z_{\odot}]=-0.63^{+0.35}_{-0.62}$; in the case of exactly zero diffuse interstellar dust, DGSAT I may even have solar metallicity. With VCC 1287 and several Coma UDGs, a general picture is emerging where cluster UDGs may be ``failed" galaxies, but the field UDG DGSAT I seems more consistent with a stellar feedback-induced expansion scenario. In the future, our approach can be applied to a large and diverse sample of UDGs down to faint surface brightness limits, with the goal of constraining their stellar ages, stellar metallicities, and circumstellar and diffuse interstellar dust content.
\end{abstract}


\maketitle

\section{Introduction}\label{sec:intro}
Ultra-diffuse galaxies (UDGs) were defined in the Coma cluster by \citet{vandokkum15a} to be exceptionally large with low optical surface brightnesses.\footnote{We emphasize that low surface brightness galaxies more generally (both small and large) have been known to exist for decades \citep[e.g.,][]{disney76,sandage84,impey88,bothun91,mcgaugh94,dalcanton97,conselice03}.} Large numbers of UDG-like objects are now being investigated across a range of environments, including groups, the outskirts of clusters, and even the field \citep[e.g.,][]{martinezdelgado16,vanderburg16,vanderburg17,roman17a}. The optical surface brightness profiles of many UDGs appear to be well described by relatively shallow central light profiles \citep[S{\'e}rsic index $n\sim1$; e.g.,][]{vandokkum15a,koda15,yagi16,martinezdelgado16,beasley16}, with their projected axis ratios suggesting a generally spheroidal structure \citep{burkert17}. Recently, it has become clear that objects which satisfy UDG-like selection criteria in terms of surface brightness and size also span a range of optical colors. In particular, there exist so-called ``blue UDGs" \citep[][]{roman17a,roman17b,shi17,greco17b,leisman17}, although it is not yet clear whether these are simply very low surface brightness dIrrs and how they might relate to red UDGs \citep[][]{trujillo17,bellazzini17,papastergis17}.

Although the formation of UDGs remains poorly understood, a few different scenarios have been proposed and it is likely that UDGs have multiple formation channels. One intriguing possibility is that the UDGs with the largest sizes and dark matter halo masses, but dwarf-like stellar masses, are ``failed" galaxies \citep{vandokkum15a,vandokkum16,yozin15}. In this picture, red UDGs in high density environments lost their gas content after forming their first few generations of stars. Alternatively, assuming that the angular momentum of the dark matter halo is what sets the sizes of galaxies, \citet{amorisco16} argued that UDGs simply represent the high-spin tail of the normal dwarf galaxy population \citep[see also][]{yozin15}. In contrast, \citet{dicintio17} proposed that the large sizes of UDGs (particularly those in relatively low density environments) can arise from stellar feedback-driven outflows and radial stellar migration, without appealing to high halo spin. \citet{rong17} presented a hybrid formation scenario, where UDGs are a mixture of high-spin dwarfs as well as objects that continued to form stars until relatively late times. \citet{greco17a} suggested, on the basis of extremely low surface brightness imaging, that some UDGs could actually be tidal debris associated with galaxy mergers while others could be tidally disrupted dwarfs \citep[see also][]{merritt16}. Finally, \citet{peng16} proposed an exotic scenario in which star formation in UDGs was rapidly truncated due to the formation of globular cluster (GC) progenitors.  

On the observational side, one concrete way to distinguish between the proposed formation scenarios is to measure the ages and metallicities of UDGs. Ages and metallicities can place direct limits on how long ago these systems formed and what their chemical enrichment histories were like. Joint age--metallicity constraints can help test the scenario of whether UDGs formed late \citep{rong17} or whether they contain only a few generations of stars from the early Universe \citep{vandokkum15a,yozin15}. More importantly, placing UDGs on the stellar mass--metallicity relation for normal dwarfs \citep[e.g.,][]{kirby13} provides a novel way of testing whether UDGs are simply a continuous extension of normal dwarfs, which has been a very difficult question to address so far \citep[see also][]{amorisco16,rong17,gu17}. 

Given the very low mean surface brightnesses of UDGs \citep[$\sim24-27$ mag arcsec$^{-2}$ within one effective radius at optical wavelengths; e.g.,][]{vandokkum15a,mihos15,yagi16}, it is quite challenging to obtain sufficiently high quality spectra for stellar population analysis. Two impressive recent exceptions are \citet{kadowaki17} and \citet{gu17}. \citet{kadowaki17} obtained spectra of four red Coma UDGs, and then visually compared the stacked spectrum (which revealed Balmer and Ca H\&K absorption lines) to a library of four simple stellar population (SSP) templates, concluding that the stacked UDG spectrum was broadly consistent with an old metal-poor stellar population. \citet{gu17} performed full (optical) spectral fitting of three Coma UDGs, and found that their galaxies were consistent with low stellar metallicities and old ages. These expensive spectroscopic results support and refine earlier studies that used only optical photometry (typically only one or two optical colors, with a very limited wavelength baseline) to broadly constrain the ages and metallicities of UDGs. For example, \citet{vandokkum15a} originally suggested that the average $g-i$ color of their sample of Coma UDGs could be reproduced by either an old metal-poor stellar population, or by one with a relatively younger age and higher (but still sub-solar) metallicity. \citet{roman17b} essentially did a similar analysis but used two optical colors instead of one ($g-r$ and $g-i$), and recovered the classical age--metallicity degeneracy \citep[i.e., they were unable to rule out a relatively young and solar metallicity stellar population; see also][]{vanderburg16}. 

In this paper, we carry out fully Bayesian Markov chain Monte Carlo (MCMC) based fitting of the optical--NIR spectral energy distributions (SEDs) of two red UDGs that do not show clear evidence of ongoing star formation (similar to red Coma UDGs), with the goal of constraining their stellar ages and metallicities. One of our UDGs is in a relatively low density environment whereas the other is in a cluster; this allows us to explore the possible role of the environment on the stellar populations of UDGs. Our rigorous SED modeling framework forces us to be explicit about our prior assumptions and allows us to marginalize over physical properties that simply cannot be constrained by the currently available photometric data. By combining new $Spitzer$-IRAC imaging with existing archival optical imaging, we will be able to break the age--metallicity degeneracy \citep[e.g.,][]{worthey94,bell00,galaz02}. Furthermore, we will directly take into account two major unknowns about UDGs: (1) whether they are consistent with a single SSP that formed in one burst or whether they had a more extended star formation history (SFH), and (2) the internal dust content of UDGs, since it is well known that age and metallicity are also severely degenerate with reddening by diffuse interstellar dust \citep[e.g.,][]{bell01}. 

This paper is organized as follows. In \autoref{sec:sample}, we describe our UDG sample and the imaging we use to construct their SEDs. In \autoref{sec:analysis}, we explain our analysis, which involves tasks related to the photometry as well as the SED fitting itself. In \autoref{sec:results}, we present our results, and then we present a discussion in \autoref{sec:discussion}. We summarize in \autoref{sec:summary}. Throughout this paper, we assume a \citet{planck14} cosmology, with $H_0 = 67.8$ km s$^{-1}$ Mpc$^{-1}$, $\Omega_m=0.307$, and $\Omega_{\Lambda}=0.693$.

\section{Sample and Data}\label{sec:sample}
Here we describe our sample of galaxies and the observational data.

\subsection{Galaxy Sample}\label{sec:galaxies}
In this paper, we focus on two UDGs (DGSAT I and VCC 1287) that are similar to red Coma UDGs in the sense that they are both optically red and do not show clear evidence of ongoing star formation \citep{vandokkum15a}. DGSAT I and VCC 1287 represent two of seven UDGs that were targeted for $Spitzer$-IRAC NIR imaging as part of Program ID 13125. These two UDGs live in quite different environments and neither of them are in the Coma cluster. \citet{martinezdelgado16} discovered DGSAT I and spectroscopically confirmed that it lives within a filament of the Pisces--Perseus supercluster ($z_{\rm red}=0.0185$), $\sim2$ Mpc in projection away from a cluster. Thus, DGSAT I is in a low density field-like environment compared to other red UDGs, which are generally found within or near clusters \citep[e.g.,][]{vanderburg17}. VCC 1287 is present in the Virgo cluster catalog of \citet{binggeli85} and was noted to be an exceptionally extended low surface brightness galaxy. \citet{beasley16} measured the spectroscopic redshift of the nucleus of VCC 1287 and found that it was similar to the mean radial velocity of Virgo ($z_{\rm red}=0.0036$). 

For a ``differential comparison" with a normal dwarf, we choose the Virgo dE VCC 1122. This galaxy was chosen simply because it already has data available in similar bandpasses as our UDGs (most importantly, IRAC channels 1 and 2). We adopt its spectroscopic redshift from \citet{toloba14} as $z_{\rm red}=0.0016$. 

\textbf{For the purpose of determining total stellar masses, we assume the luminosity distances to DGSAT I (83 comoving Mpc) and VCC 1287 (16 comoving Mpc), which are in relatively good agreement with \citet{martinezdelgado16} and \citet{beasley16}, respectively. For the dE VCC 1122, we assume a distance of 16 comoving Mpc, which is consistent with the Virgo cluster.} 

\subsection{$Spitzer$-IRAC NIR Imaging}\label{sec:spitzer}
Here we describe our new $Spitzer$-IRAC observations. DGSAT I was observed with $Spitzer$-IRAC \citep{fazio04,werner04} at 3.6 $\mu$m (IRAC1) and 4.5 $\mu$m (IRAC2) on 2016 October 23, whereas VCC 1287 was observed only at 3.6 $\mu$m on 2017 April 13 (archival 4.5 $\mu$m imaging will be discussed below). We used dithered observations for both galaxies (observation IDs 61004544 and 61005312 in Program ID 13125 for DGSAT I and VCC 1287, respectively). Only one pointing per galaxy was needed because the UDG sizes are much smaller than the IRAC field of view ($\sim5\times5$ arcmin$^2$). We used 50 medium-scale cycling dithers (with $\sim65$ arcsec median dither separation) to eliminate array-dependent or transient artifacts (e.g., bad pixels, radiation hits, residual images, scattered light). The frame times used were 93.6 sec and 96.8 sec in IRAC1 and IRAC2, respectively; these are the longest allowed frame times and are recommended for faint object observations. For each UDG, we therefore spent 4680 sec in IRAC1 and 4840 sec in IRAC2. Note that the IRAC native pixel scale is $\sim1.2$ arcsec/px in channels 1 and 2; the mosaics were resampled to $0.6$ arcsec/px as is standard practice. Thus, the IRAC images have significantly worse spatial resolution than our archival optical images (discussed below in \autoref{sec:archival}).  

For IRAC 4.5 $\mu$m observations of VCC 1287, we used archival data from the $Spitzer$ Heritage Archive (SHA) to produce a mosaic. We used data from Program ID 10015, which had 12 separate observations (observation IDs 50299392, 50300160, 50300928, 50301440, 50301952, 50302464, 50302976, 50303488, 50304000, 50382592, 50382848, and 50383104). Each of those 12 observations used nine medium-scale cycling dithers with 30 sec frame time (26.8 sec of exposure time per frame). This resulted in 2894.4 sec of time on the source, although the whole extent of VCC 1287 was not completely covered in a few frames. These observations were unfortunately taken at a time (September 2014) when IRAC data suffered from striping (randomly varying high and low intensity horizontal rows). The individual data frames produced by version S19.2.0 of the IRAC pipeline were ``destriped'' using a custom-built IDL routine (James Ingalls, private communication). In addition, the ``column pulldown effect" was corrected using the \texttt{imclean} code.\footnote{The ``column pulldown effect" is explained in section 7.2.4 of the IRAC Instrument Handbook. The \texttt{imclean} code is publicly available on the $Spitzer$ website at \url{http://irsa.ipac.caltech.edu/data/SPITZER/docs/dataanalysistools/tools/contributed/irac/imclean/}.} Finally, the individual frames were mosaicked together using the default parameters in the $Spitzer$ Mosaicker and Point Source Extractor (MOPEX) software.\footnote{The $Spitzer$ \texttt{MOPEX} software is publicly available at \url{http://irsa.ipac.caltech.edu/data/SPITZER/docs/dataanalysistools/tools/mopex/}.}

For the dE VCC 1122, we use IRAC1 and IRAC2 mosaics from $Spitzer$ Program ID 60173 \citep[see][for details]{krick11}. The data were taken with a similar setup as for our new observations above. We also use archival IRAC 8.0 $\mu$m (IRAC4) imaging centered on this dE from $Spitzer$ Program ID 20606 (PI: Bressan). For the IRAC4 images, there were 35 individual frames taken with an exposure time of 26.8 sec per frame.

\subsection{Archival Optical Imaging}\label{sec:archival}
Here we briefly describe the archival optical imaging that we use for each galaxy. For DGSAT I, we use the same archival Subaru Suprime-Cam imaging that was presented in \citet{martinezdelgado16}. We adopt their calibrated images directly, but perform the background subtraction, masking of contaminants, and photometry ourselves; the Vega magnitude zeropoints for the Johnson-Cousins $V$ and $I$ bands from \citet{martinezdelgado16} are 33.95 mag and 33.35 mag, respectively. We transform these $V$ and $I$ Vega magnitudes to $V$ and $I$ AB magnitudes using the linear offsets given in Table 1 of \citet{blanton07}. Specifically, $V_{\rm AB} = V_{\rm Vega}+0.02$ mag and $I_{\rm AB} = I_{\rm Vega}+0.45$ mag. This means that $V-I\approx1.0$ Vega mag for \citet{martinezdelgado16} would correspond to $V-I\approx0.6$ AB mag for us (their Figure 7 suggests that the average global $V-I$ color of DGSAT I is $\approx0.8$ Vega mag, after MW reddening corrections). 

For VCC 1287, we originally tried to use the public archival Canada--France--Hawaii Telescope (CFHT) MegaCam $u^*griz$ imaging, as was done by \citet{beasley16}. However, those images are already background-subtracted in a way that is not optimal for extended low surface brightness galaxies \citep{gwyn08}, and the background maps themselves were not saved. The background mesh size was smaller than the UDG, which means that a significant fraction of the UDG light was considered to be part of the background and thus subtracted off. Instead, in this work we use proprietary CFHT-MegaCam $u^*giz$ imaging ($r$-band is unavailable) that is based on the same data but processed using the more appropriate ``Elixir-LSB" NGVS pipeline \citep[see][]{ferrarese12,duc15}. The UDG appears $\sim3-4\times$ brighter and $\sim50\%$ more extended in this proprietary Elixir-LSB-processed CFHT imaging compared to the public archival CFHT imaging; we discuss this further in \autoref{sec:VCC 1287} and Appendix \ref{sec:elixir}. 

For the Virgo dE VCC 1122, we use public archival SDSS $ugriz$ imaging \citep[DR14;][]{abolfathi17}.

\begin{figure*} 
\begin{center}
\includegraphics[width=\hsize]{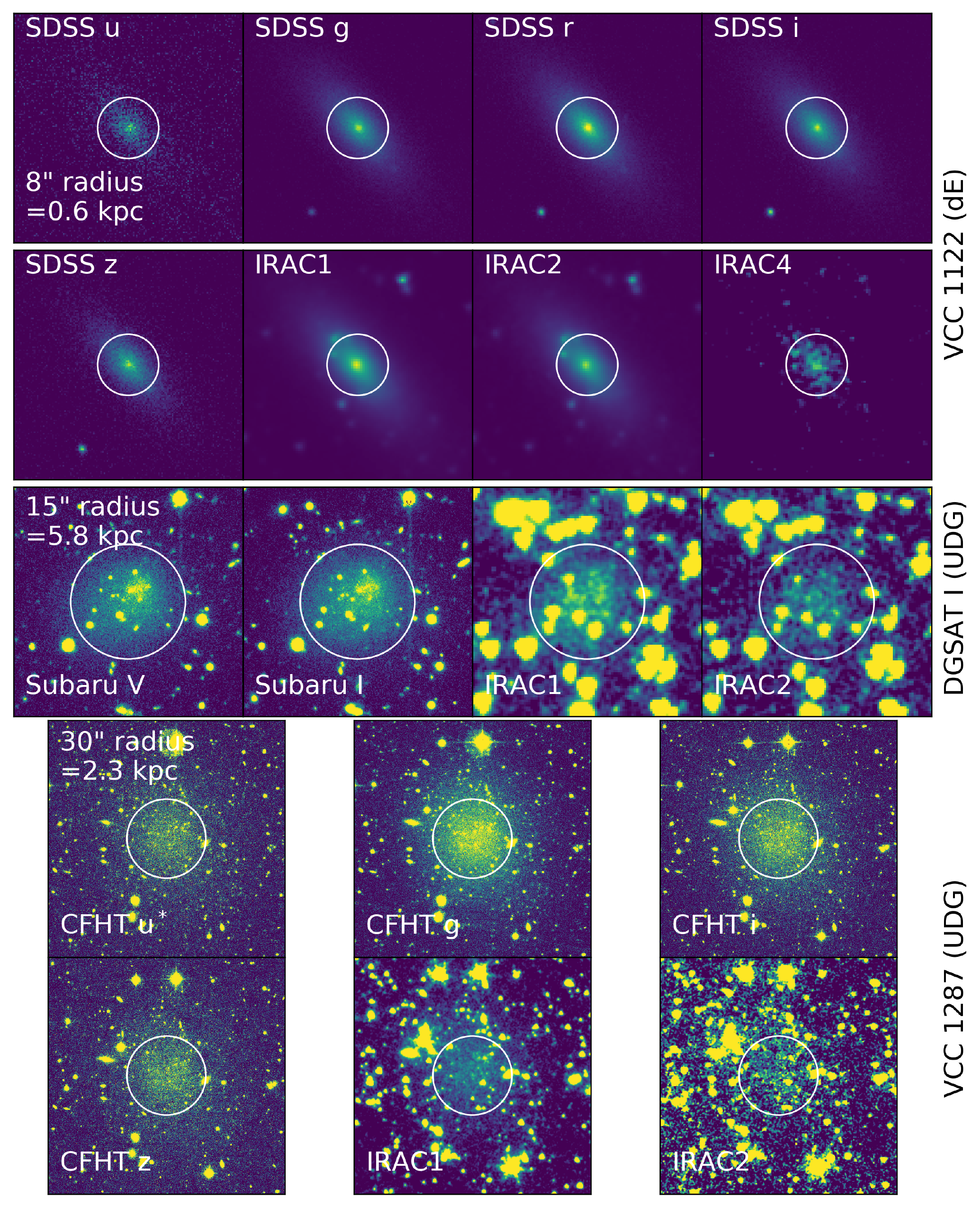}
\end{center}
\caption{Postage stamps in each of the bandpasses analyzed for VCC 1122, DGSAT I, and VCC 1287. The sizes of the cutouts are $1\times1$ arcmin$^2$ for VCC 1122, $1\times1$ arcmin$^2$ for DGSAT I, and $3\times3$ arcmin$^2$ for VCC 1287. The white circles show the photometry apertures (8 arcsec radius for VCC 1122, 15 arcsec radius for DGSAT I, and 30 arcsec radius for VCC 1287)\textbf{; the corresponding physical aperture sizes in kpc are also written}. In all images of DGSAT I and VCC 1287, north is up and east is left; in the images for the dE, north is right and east is up (maintaining SDSS convention since the dE is near the edge of the SDSS FOV). All images shown are background-subtracted, and the colorbar limits have been arbitrarily adjusted for each subplot to highlight the galaxy against background residuals. Although quite faint, especially in the IRAC imaging, both UDGs are significantly detected in all bands (see text).}
\label{fig:stamps}
\end{figure*}

\section{Analysis}\label{sec:analysis}
Now we turn to the analysis, which involves masking contaminants, subtracting the background, aperture photometry, and then SED fitting. Because color measurements of low surface brightness galaxies are extremely challenging, we also devote a subsection to measurements of the statistical and systematic errors in each dataset.

\subsection{Masking and Background Subtraction}\label{sec:maskbkg}
Foreground and background objects, regardless of whether they are point or extended sources, can artificially increase the measured flux of UDGs. We use Source Extractor \citep{bertin96} to create bad pixel masks of detected objects in a semi-automated fashion. The parameters of Source Extractor were iteratively tuned by hand until all non-smooth features superimposed on each UDG were detected and therefore masked. We allowed the Source Extractor parameters to vary in different bandpasses, but ensured through visual inspection that the masks for different optical bands were roughly similar. For $Spitzer$-IRAC imaging in particular, more aggressive deblending and a lower detection threshold was needed owing to the worse spatial resolution and the increased contribution of high-redshift NIR background sources. 

In the Subaru $VI$ optical imaging of DGSAT I, there is an off-center overdensity and it is not clear whether it is associated with the UDG or is instead a conglomeration of contaminant sources \citep[see][]{martinezdelgado16}. We cannot robustly constrain the stellar population of the overdensity with our limited broadband photometric data. In particular, Figure 7 of \citet{martinezdelgado16} shows that the bluest part of this overdensity would be enclosed within a very small ellipse with $a=1.4"$ and $b=0.6"$ (semi-major and semi-minor axis lengths, respectively). Given the poor spatial resolution of IRAC (native pixel scale of 1.2 arcsec/px), we cannot reliably measure the NIR flux within such a small aperture (significant and highly uncertain aperture corrections would be needed). Therefore, we mask out the overdensity region in all bands using a circular aperture of radius 3 arcsec. We find that the overdensity region accounts for $\sim15\%$ of the total galaxy flux within a 15 arcsec aperture in all bands, after masking out other contaminants.

In the optical imaging of DGSAT I, a spatially smooth and relatively uniform background was subtracted off (in the $V$-band, the median background surface brightness level was $\sim19.7$ mag arcsec$^{-2}$). The global background levels were good at the $\sim3\%$ fractional flux level (i.e., the average ratio of background RMS map to background map itself is $\sim3\%$). For the VCC 1287 CFHT optical $u^*giz$ imaging, the background was already subtracted off by the Elixir-LSB pipeline \citep[see][]{ferrarese12,duc15}. However, we found that there remained an overall non-zero median background level that was $\sim0.5$ counts/px in $u^*gi$ and $\sim2$ counts/px in $z$. We used Source Extractor to subtract off this median overall background offset so that the median background level was $\sim0$ counts/px in every optical image. We made sure to use a background mesh size that was much larger than the UDG to prevent subtraction and distortion of galaxy light. If we forego this correction, then the $u-z$ color is inexplicably red and the $z-[3.6]$ color is incredibly blue, suggesting that our median background subtraction is warranted. For the SDSS $ugriz$ imaging of dE VCC 1122, the background was already adequately subtracted by the SDSS pipeline and its uncertainties are negligible given how bright the dE is. 

The background subtraction for $Spitzer$-IRAC was more involved because we had to do the ``first-frame correction" (to address imperfect bias subtraction; see section 5.1.10 of the IRAC Instrument Handbook). Specifically, we rectified each individual data frame for history effects in the IRAC arrays in two steps. First, a per-pixel correction was made according to the amount of time an array idles before the start of an integration. The corrections were derived from 6000 single frame exposures in each array in the IRAC sky dark field. The characteristics of the first-frame effect have changed a little since they were first calibrated, and the best results were obtained when we scaled up the IRAC1 correction images by 10\%. In IRAC1, the correction images have considerable spatial structure, while in IRAC2 they are nearly flat except for seven columns. As applied to these data, in IRAC1, the RMS correction for all pixels was $\sim4$ kJy sr$^{-1}$ with a standard deviation of 1.2 kJy sr$^{-1}$ (i.e., the spread in corrections to any pixel, relative to the spatial means). In IRAC2, both values were $\sim1$ kJy sr$^{-1}$. The typical correction is much smaller than the read noise, and the uncertainties in the corrections are at least a factor of five smaller. Therefore, we do not include any systematic magnitude errors due to these first-frame corrections.

In the second step for the IRAC background subtraction, the mean background in each image is computed, and then a function is fitted to the means and subtracted from the images. In a sequence of 100 sec frames such as ours, the backgrounds undergo a rapid drop and then a slow increase. The mean backgrounds as a function of time are well fitted by a function of the form
\begin{equation}
y = c_0 + \sum_{i=1}^{n\leq3} c_i e^{-r_i t} \;,
\end{equation}
where $t$ is the time since the beginning of the third frame. The initial drop was not fitted in the DGSAT I data; two exponential terms sufficed in IRAC1 and one in IRAC2. For VCC 1287 IRAC1, the initial drop was included in the fit along with two exponential terms. The uncertainties in these first-frame effect corrections are negligible compared to other sources of systematic error. Because the VCC 1287 archival IRAC2 data suffered from fairly severe striping, we did not make the first-frame effect corrections for those data as their noise is completely dominated by the residuals from the striping correction. Instead, we used Source Extractor to model and subtract a smooth and relatively uniform background. Finally, since the IRAC imaging of the Virgo dE VCC 1122 was subject to first-frame effect corrections done differently than here \citep[see][]{krick11}, we also ran Source Extractor to subtract the background in those images.

\subsection{Aperture Photometry}\label{sec:photometry}
Our photometry is based on circular apertures with fixed radius across all bandpasses for a given object. The masks described in the previous section are used to ignore any contaminated pixels when computing the summed aperture counts. The apertures were chosen to roughly correspond to the optical half-light radius given in the literature and to avoid significant nearby contamination in the IRAC imaging. The centroids of the apertures correspond to the RA and Dec given in \citet{martinezdelgado16} for DGSAT I, \citet{beasley16} for VCC 1287 and \citet{toloba14} for VCC 1122. For DGSAT I, we use a radius of 15 arcsec \citep[this is the S{\'e}rsic optical half-light radius given by][]{martinezdelgado16}. For VCC 1287, we use a 30 arcsec radius \citep[following][]{beasley16}.\footnote{In Appendix \ref{sec:galfit}, we show that when we run GALFIT on the new Elixir-LSB CFHT imaging, we find that VCC 1287 is $50\%$ more extended than what \citet{beasley16} found. Specifically, we find $R_e=46.4$ arcsec. For our fiducial analysis, we will continue to adopt a 30 arcsec radius so that we can later give a revised estimate of the dynamical mass to stellar mass ratio, since \citet{beasley16} conveniently provide a dynamical mass measurement within $\sim30$ arcsec.} For the dE VCC 1122, we assume a radius of 8 arcsec to avoid contamination from nearby point sources in the IRAC imaging. Our aperture radius for the dE is a factor of $\sim2$ smaller than the $r$-band half-light radius given in Table 4 of \citet{toloba14}, but it encloses the brightest part of the dE (the central spheroidal component).

We caution that our circular aperture photometry technique is rather crude and does not capture ``all" of the light from a galaxy. However, the main effect of using model-based magnitudes, which come from integrating a best-fit structural model out to infinity, is to change the overall normalization of the best-fit SED (i.e., the stellar mass or luminosity in different bandpasses). Assuming no large-scale spatial color gradients, the fact that our apertures are fixed across significantly different wavelengths means that our colors and SED shapes will be self-consistently measured for each object. This accordingly allows us to more confidently infer the average global stellar population properties of a galaxy than if there were significant systematic offsets in different bandpasses due to the underlying structural models. We attempted to use GALFIT \citep{peng02,peng10} in two different ways: letting all S{\'e}rsic parameters vary in every bandpass, or fixing the profile in all bands to match that of the band with the highest S/N. While the formal statistical errors on the model parameters in each bandpass were rather small in some cases, systematic uncertainties were evident when visually inspecting the model residuals (e.g., optical surface brightness profiles not fully capturing NIR light due to variations needed in the S{\'e}rsic index, radius, axis ratio and/or PA). Thus, we settled on the cleaner and more straightforward method of fixed-size circular apertures chosen to encapsulate most of the galaxy light in all bandpasses. 

The AB magnitudes resulting from our aperture photometry in each bandpass are given in \autoref{tab:photometry1122} for VCC 1122, \autoref{tab:photometryDGSAT I} for DGSAT I and \autoref{tab:photometry1287} for VCC 1287. The magnitudes have been corrected for foreground extinction due to dust within the Milky Way (MW) assuming the values in the relevant bandpasses from \citet{schlafly11}.\footnote{We made use of \url{http://irsa.ipac.caltech.edu/applications/DUST/}.}. In the case of IRAC, we also multiply the measured fluxes by the recommended aperture correction factors.\footnote{These aperture correction factors are given by equation (4.20) and Table 4.8 of the IRAC Instrument Handbook.} We do not apply any aperture corrections for the ground-based optical imaging.

\subsection{Photometric Uncertainties}\label{sec:photerr} 
We consider purely statistical uncertainties on our measured magnitudes as well as several sources of systematic error. For the statistical uncertainties, our default approach was to calculate the formal S/N using the aperture-summed electrons from the background-subtracted source ($e_s$) as well as from the background itself ($e_b$): 
\begin{equation}\label{eq:snr}
\frac{S}{N} = \frac{e_s}{\sqrt{e_s + e_b}}\,.
\end{equation}
Read noise was neglected except for SDSS, where it is a larger source of error than the Poisson sky noise.\footnote{For CFHT optical imaging, the images are already background subtracted and the original background map was not saved. However, the image headers have the minimum and maximum original sky levels recorded. We therefore assume that the original background level per pixel across an entire image is characterized by this average sky level. For the SDSS read noise, see \url{https://data.sdss.org/datamodel/files/BOSS_PHOTOOBJ/frames/RERUN/RUN/CAMCOL/frame.html}.} The resulting aperture-summed S/N values in each band are given in \autoref{tab:photometry1122}, \autoref{tab:photometryDGSAT I}, and \autoref{tab:photometry1287} for VCC 1122, DGSAT I, and VCC 1287, respectively. For comparison, we also list the median S/N per pixel, which is computed using the same equation. It is clear that while in many bandpasses the UDGs are ``in the noise" in terms of their surface brightness, the UDGs are significantly detected in every bandpass based on their aperture-summed counts. To get the actual statistical uncertainty on the aperture magnitude, we first convert the aperture-summed S/N into a formal fractional flux uncertainty via $\Delta f = 1/(S/N)$, and then into a magnitude error via $2.5\log_{10}(1+\Delta f)$, following common practice \citep[e.g.,][]{kniazev04}. This calculation resulted in statistical errors at the $\lesssim0.01$ AB mag level in every bandpass for all three objects. We recovered similarly small statistical magnitude errors using a Monte Carlo approach where Gaussian random noise was added to each pixel (with zero mean and standard deviation given by the background RMS level for that pixel), the process was repeated 100 times, and the standard deviation of the resulting magnitudes were taken.

We accounted for four sources of systematic error: masking, overall background subtraction bias, photometric calibration offset, and IRAC-specific corrections. These errors were summed in quadrature together with the statistical uncertainties derived above. In \autoref{sec:maskbkg} we mentioned that the background maps we derived ourselves were good at the $\lesssim3\%$ level (based on the average ratio of the background RMS to the background level itself). As an alternative way to measure the errors due to masking and overall background subtraction bias, we re-did our aperture measurements 100 times, each time randomly toggling 50\% of the pixels within the aperture to the opposite mask value (either masked or unmasked). We simultaneously also added to every pixel a Gaussian random value based on the background map RMS level in that pixel (with zero mean). The standard deviation of the resulting aperture fluxes were typically at the $\sim0.01$ AB mag level. Instead of using such low errors, we conservatively budgeted 5\% fractional flux error ($\sim0.053$ mag error) due to masking and background subtraction bias. For the VCC 1287 CFHT data, we instead assumed 10\% instead of 5\% to account for our additional median background subtraction described above in \autoref{sec:maskbkg}.

The errors in the photometric calibration to get onto the AB magnitude system were conservatively assumed to be at the 3\% fractional flux uncertainty level in all optical bandpasses and 2\% for IRAC. For IRAC, we also budgeted 4\% and 2\% fractional flux uncertainty due to the integrated aperture flux correction factor and the array location-dependent color correction, respectively.\footnote{Since the array location-dependent color correction (which addresses assumptions made during the IRAC flat-fielding process) is negligible for our purposes, we do not actually apply it and instead only include it as an additional possible source of systematic error. See Section 4.5 of the IRAC Instrument Handbook: \url{http://irsa.ipac.caltech.edu/data/SPITZER/docs/irac/iracinstrumenthandbook/}.}. The final magnitude errors, which are dominated by systematics, are given in \autoref{tab:photometry1122}, \autoref{tab:photometryDGSAT I}, and \autoref{tab:photometry1287} for VCC 1122, DGSAT I, and VCC 1287, respectively. 

\begin{table}
\caption{Aperture Photometry for dE VCC 1122}
\centering
\begin{tabular}{ccccc}
\hline
Band & AB Magnitude & Magnitude Error & S/N & S/N/px \\\hline
$u$ & 17.17 & 0.06 & 131.0 & 3.5 \\
$g$ & 15.74 & 0.06 & 667.0 & 16.1 \\
$r$ & 15.11 & 0.06 & 886.6 & 21.5 \\
$i$ & 14.78 & 0.06 & 879.1 & 21.6 \\
$z$ & 14.64 & 0.06 & 459.1 & 11.4 \\
IRAC1 & 15.42 & 0.07 & 4281.1 & 155.0 \\
IRAC2 & 15.89 & 0.07 & 3290.5 & 121.5 \\
IRAC4 & 16.82 & 0.07 & 406.7 & 14.2 \\
\end{tabular}
\tablecomments{The photometry is based on an 8 arcsec circular aperture centered on the Virgo dE VCC 1122. The IRAC magnitudes include a standard aperture correction, but no such aperture corrections were made for the optical data. The uncertainties include purely statistical uncertainties as well as several sources of systematic errors (see text). The $ugriz$ bands are based on SDSS filters. The magnitudes include corrections for foreground MW dust attenuation. Both the S/N integrated within the aperture as well as the median S/N per pixel within the aperture are given.}
\label{tab:photometry1122}
\end{table}

\begin{table}
\caption{Aperture Photometry for UDG DGSAT I}
\centering
\begin{tabular}{ccccc}
\hline
Band & AB Magnitude & Magnitude Error & S/N & S/N/px  \\\hline
$V$ & 18.87 & 0.08 & 88.9 & 0.6 \\
$I$ & 18.55 & 0.08 & 110.5 & 0.8 \\
IRAC1 & 19.12 & 0.07 & 348.6 & 7.9 \\
IRAC2 & 19.43 & 0.08 & 137.2 & 3.1 \\
\end{tabular}
\tablecomments{Similar to \autoref{tab:photometry1122} but for photometry of the field UDG DGSAT I within a 15 arcsec circular aperture. The $V$ and $I$ bands are based on Subaru-$Suprimecam$ Johnson-Cousins filters.}
\label{tab:photometryDGSAT I}
\end{table}

\begin{table}
\caption{Aperture Photometry for UDG VCC 1287}
\centering
\begin{tabular}{ccccc}
\hline
Band & AB Magnitude & Magnitude Error & S/N & S/N/px \\\hline
$u^*$ & 18.21 & 0.11 & 43.1 & 0.2 \\
$g$ & 17.07 & 0.11 & 31.5 & 0.1 \\
$i$ & 16.35 & 0.11 & 55.4 & 0.2 \\
$z$ & 16.20 & 0.11 & 60.8 & 0.2 \\
IRAC1 & 17.42 & 0.07 & 812.1 & 9.7 \\
IRAC2 & 17.72 & 0.07 & 283.9 & 3.6 \\
\end{tabular}
\tablecomments{Similar to \autoref{tab:photometry1122} but for photometry of the Virgo UDG VCC 1287 within a 30 arcsec circular aperture. The $u^*giz$ bands are based on CFHT-MegaCam filters.}
\label{tab:photometry1287}
\end{table}

\subsection{SED Fitting with \texttt{prospector}}\label{sec:prospector}
With our aperture photometry in hand, we run the fully Bayesian MCMC-based \texttt{prospector} inference code \citep[][and B. Johnson, in preparation]{leja17} on the resulting SEDs. Because models are generated on the fly, the \texttt{prospector} code allows for flexible model specification with larger numbers of parameters than are computationally tractable in typical grid-based searches. However, as in grid-based inference, it is still necessary to fully account for (numerous) degeneracies in the stellar population parameters, something that is difficult to accomplish in techniques based on optimization. This is accomplished in \texttt{prospector} through MCMC sampling of the Bayesian posterior probability distribution.

The flexibility of \texttt{prospector} is aided by the tight coupling with the Flexible Stellar Population Synthesis package \citep[FSPS;][]{fsps1,fsps2,fsps3}, where numerous parameters affecting the SED can be varied; all of the FSPS parameters are potentially free parameters in \texttt{prospector}. For our particular \texttt{prospector} setup, we fix most parameters to some value and adopt the following default models and prior assumptions. We fix the redshifts of the three galaxies to the values given in \autoref{sec:galaxies}. We use the MILeS stellar spectral library \citep{sanchezblazquez06,falconbarroso11}. We also adopt the Padova isochrones \citep{marigo07,marigo08}, which only allow us to explore stellar metallicities in the range $-2.0<[Z/Z_{\odot}]<0.2$. The FSPS models used in \texttt{prospector} assume solar-scaled abundances (i.e., $[\alpha/\rm Fe]=0$). However, the effects of $\alpha$-element enhancement on broadband SEDs are expected to be much smaller than other effects. We allow for emission from circumstellar dust around thermally-pulsating AGB stars using the \citet{villaume15} models, and account for diffuse interstellar dust emission using the default \citet{draine07} models. The TP-AGB models are especially important for predicting NIR fluxes (particularly at ages of $<3$ Gyr). Nebular emission lines are enabled by default according to the prescription of \citet{byler17}. Finally, we adopt the \citet{kroupa01} initial mass function, which is the default in FSPS. For MCMC, we use the \texttt{emcee} package \citep{foremanmackey13} with 128 walkers, Powell optimization, three rounds of burn-in (512 iterations each), and 3000 iterations thereafter. 

We place very strong priors on the form of the SFH and the shape of the dust attenuation curve. Specifically, we assume an exponentially declining $\tau$ model for the SFH and the \citet{calzetti00} attenuation curve for dust within the galaxies. The $\tau$ model SFH does not allow us to constrain bursty or stochastic SFHs, but it does allow for SSPs in the limit that the e-folding timescale $\tau\rightarrow0$ Gyr. We fit five free parameters: stellar mass ($M_*$), stellar metallicity ($[Z/Z_{\odot}]$), e-folding timescale ($\tau$), age since the first onset of star formation ($t_{\rm age}$), and the dust optical depth at 5000\AA\ ($\tau_{5000}$).\footnote{For DGSAT I, our fiducial setup involves fitting five free parameters despite having only four data points. Given that we are using a fully Bayesian approach with MCMC, this is not an issue (what matters is the constraining power of the data, not simply the number of data points). Indeed, as we will show later, our limited DGSAT I data do have sufficient constraining power to rule out parts of stellar population parameter space. Of course, extending the wavelength baseline would help provide better constraints.} The following linearly uniform priors were used for all five of these free parameters: $M_*=10^{6-10}M_{\odot}$, $[Z/Z_{\odot}]=-2.0$ to $0.2$, $\tau=0.1-10$ Gyr, $t_{\rm age}=0.1-14$ Gyr, and $\tau_{5000}=0-4$.\footnote{While FSPS specifically fits for the dust optical depth at 5000\AA\ as the normalization of the assumed dust attenuation curve, it is more common in the literature to report $A_V=1.086\times\tau_{5000}$. Thus, throughout this paper we report $A_V$.} In \autoref{sec:systematics}, we will discuss the impact of these assumptions on our results; in short, using different prior assumptions does not significantly change our conclusions.

\subsection{Total Stellar Masses with GALFIT}\label{sec:mstartot}
Later in this paper, we will consider the location of our objects on the stellar mass--metallicity relation for dwarf galaxies. For that, we will need total stellar masses, not aperture stellar masses. We derived total stellar masses for our objects by running GALFIT \citep{peng02,peng10} on a single bandpass image for each galaxy ($I$ for DGSAT I; $i$ for VCC 1287 and VCC 1122). For DGSAT I and VCC 1287, we assumed a single S{\'e}rsic profile and allowed all parameters to be free. VCC 1122 was more complicated and required three separate structural components (two S{\'e}rsic profiles and one exponential disk); with only a single S{\'e}rsic profile, significant residuals were leftover appearing as ``side lobes." It is well known that many dEs have complicated structural properties and require multiple components \citep[e.g.,][]{janz14}. Our GALFIT results and residuals for all three galaxies are reasonable; we defer details to Appendix \ref{sec:galfit}. 

We multiplied the total enclosed GALFIT magnitudes by the corresponding aperture stellar $M_*/L$ in the same bandpass that we found above with \texttt{prospector}. Note that we use the surviving stellar mass, not the total formed mass (which would otherwise include stellar remnants). Specifically, our aperture $M_*/L_I=1.3M_{\odot}/L_{\odot}$ for DGSAT I, $M_*/L_i=1.3M_{\odot}/L_{\odot}$ for VCC 1287, and \textbf{$M_*/L_i=1.8M_{\odot}/L_{\odot}$ for VCC 1122}. When doing this conversion, we are of course assuming that there are no spatial gradients in the stellar $M_*/L$ ratio. While this is likely fine for DGSAT I and VCC 1287, it is entirely possible that the different structural components of the dE have different stellar $M_*/L$ ratios. It is beyond the scope of this paper to do more detailed modeling of spatially varying stellar $M_*/L$ ratios for our objects. Since the formal GALFIT errors are negligible and the \texttt{prospector} errors on aperture stellar masses are also small, we conservatively assume that all of our total stellar masses have a systematic uncertainty of a factor of two. Our total stellar masses are $4.8\times10^8M_{\odot}$ for DGSAT I, $2.0\times10^8M_{\odot}$ for VCC 1287, and \textbf{$1.1\times10^9M_{\odot}$ for VCC 1122.}

\section{Results}\label{sec:results}
Here, we present our results on the stellar populations of the two UDGs DGSAT I and VCC 1287 compared to the dE VCC 1122.

\subsection{Color Comparison to Single SSPs}\label{sec:colorcolor}
\begin{figure} 
\begin{center}
\includegraphics[width=\hsize]{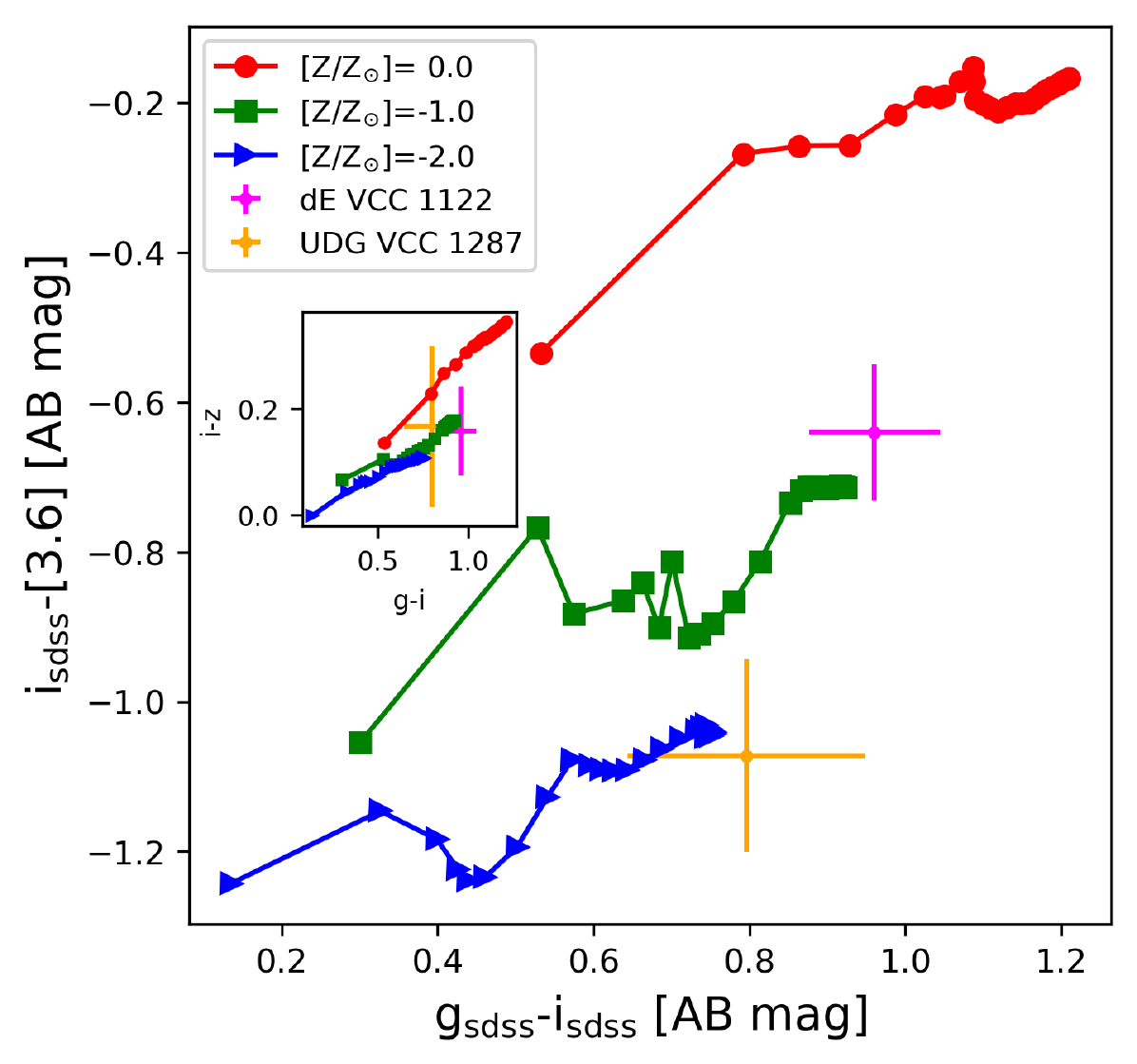}
\includegraphics[width=\hsize]{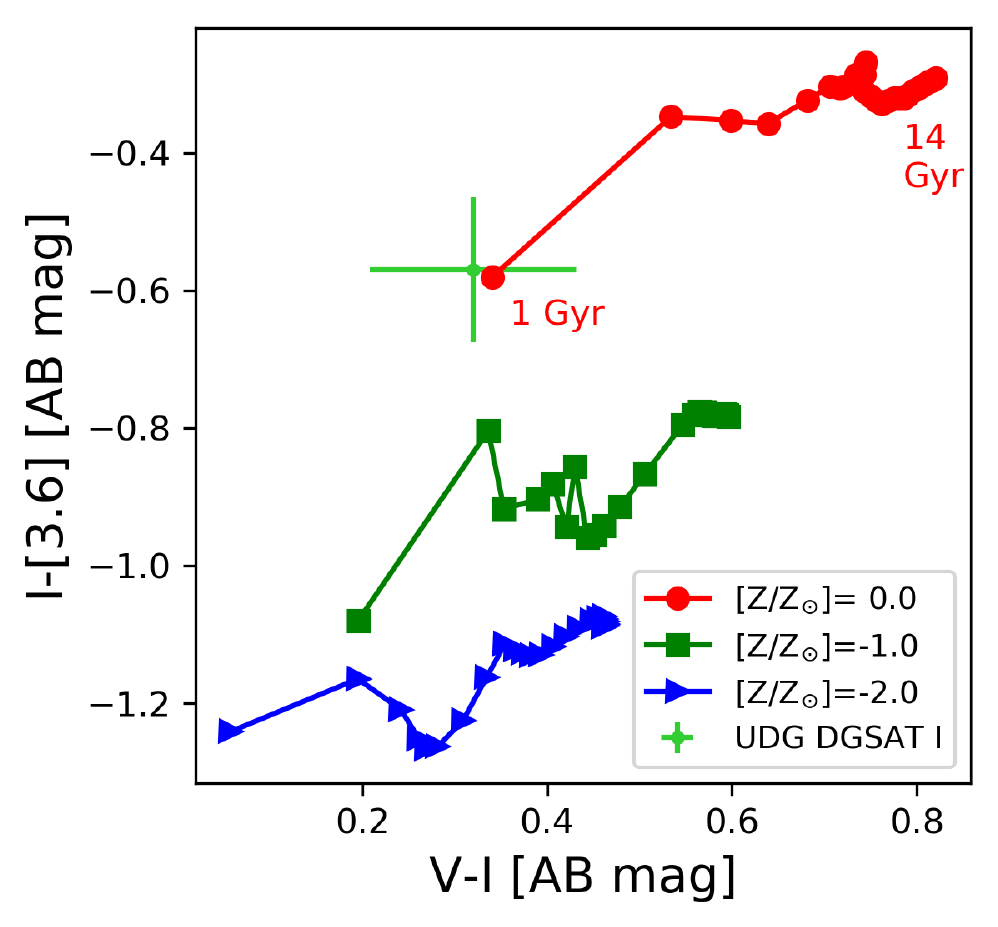}
\end{center}
\caption{Optical--NIR color--color diagrams compared to evolutionary tracks for single SSPs from FSPS with three different metallicities ($[Z/Z_{\odot}]=-2.0,-1.0,0.0$). We have transformed the CFHT $g, i$ magnitudes to SDSS magnitudes for VCC 1287. A similarly reliable transformation of Johnson-Cousins $I$ to SDSS $i$ for DGSAT I is not available, so we show DGSAT I in its own color--color diagram. The SSPs age from left to right in these diagrams, with markers representing 0.5 Gyr steps from 1 to 14 Gyr. Based on this simple color comparison, VCC 1287 (orange) is more metal-poor while DGSAT I (lime-green) is younger and more metal-rich compared to the dE VCC 1122 (magenta). The inset panel in the top subplot demonstrates that optical-only color--color diagrams do not have the dynamic range required to distinguish between SSPs with different stellar metallicities.}
\label{fig:colorcolor}
\end{figure}

Before jumping straight into the SED fitting results, it is useful to compare our observed photometry to single SSP evolutionary tracks in color--color diagrams. This is important for two reasons: (1) the observed magnitudes are independent of the models, and (2) the single SSPs are the basis for more complicated models involving SFHs and dust attenuation. In \autoref{fig:colorcolor}, we show where the UDGs DGSAT I and VCC 1287 and our comparison dE VCC 1122 fall in optical--NIR color--color diagrams. A filter transformation from CFHT $gi$ to SDSS $gi$ is straightforward for VCC 1287, but a similarly reliable conversion of Johnson-Cousins $I$ to SDSS $i$ or CFHT $i$ is not available. Therefore, we show the Virgo objects VCC 1287 and VCC 1122 in the same diagram but DGSAT I separately. 

One can already see from \autoref{fig:colorcolor} that the optical--NIR color is a powerful discriminator between single SSPs of different metallicities; such wide and clear separation in terms of metallicities is not possible using optical colors alone \citep[see also][]{laine16}. The dE VCC 1122 is consistent with an old, intermediate sub-solar metallicity SSP. DGSAT I appears to be rather young, with solar metallicity. In contrast, VCC 1287 is consistent with a very old and very metal-poor single SSP track. This already suggests that both UDGs have significantly different stellar populations compared to our comparison dE. 

Of course, this type of visual color--color comparison is very simplistic, and in reality there are a myriad of degeneracies in the stellar population models that need to be marginalized over. That is exactly the point of \texttt{prospector}, which we turn to next.

\subsection{dE VCC 1122}\label{sec:VCC 1122}
We begin with our comparison dE VCC 1122, for which we have two fitting scenarios: (1) we exclude IRAC4 from the fit and assume a uniform prior over optical attenuation $A_V=0-4$ mag, and (2) we include IRAC4 in the fit with the same uniform prior on $A_V$. In \autoref{fig:seds}, we show the highest likelihood model spectrum from both of these scenarios, compared with the observed photometry. For completeness, we plot the $16-84$ percentile spread in predicted flux density at each wavelength from the first fitting scenario without IRAC4. One can see that while the highest likelihood model spectra in both scenarios fit the optical--NIR data well, the predictions diverge dramatically at longer wavelengths. In other words, there is a severe degeneracy with both diffuse interstellar dust as well as circumstellar dust emission that our baseline optical--NIR data alone are incapable of constraining. However, we find that including IRAC4 in the fit rules out a large part of parameter space with strong mid-IR dust emission features from TP-AGB and/or post-AGB stars. 

The age--metallicity degeneracy is broken, but the dust--metallicity degeneracy is strong, as shown by the significantly different marginalized posteriors for stellar metallicity in the two dust scenarios (\autoref{fig:posteriors}). Without IRAC4, the metallicity posterior for the dE is consistent with either intermediate sub-solar metallicity ($[Z/Z_{\odot}]\approx-0.7$) or very low metallicity ($[Z/Z_{\odot}]\approx-1.8$). This happens because dust is an additional source of reddening beyond the stellar metallicity, and so a higher amount of dust means that a lower metallicity is needed to reproduce the same observed red color (at a fixed age). After elevated mid-IR dust emission models are strongly ruled out with IRAC4, the metallicity posterior becomes unimodal and peaks at an intermediate sub-solar metallicity, which is reasonable for normal dwarfs.

With the IRAC4 data point included in the fit, the dE VCC 1122 appears to be old \textbf{($\gtrsim8.6$ Gyr)} with intermediate sub-solar metallicity ($[Z/Z_{\odot}]\approx-0.7$). This is roughly consistent with the location of this dE in the optical--NIR color--color diagram (\autoref{fig:colorcolor}) compared to single SSP evolutionary tracks, even though we have now marginalized over the effects of dust and extended SFHs. In \autoref{tab:freeparams} and \autoref{tab:freeparamsnodust}, we give summary statistics for the free parameters in the two dust scenarios. 

We note that in the literature there is no consensus on the age and metallicity of VCC 1122. Results differ based on observing and fitting methodology, stellar population models assumed, bandpass coverage, and spatial regions probed. \citet{toloba14} found that VCC 1122 is $3.1^{+0.5}_{-0.4}$ Gyr old with $[Z/Z_{\odot}]=-0.4\pm0.1$ (within their $R_e\approx17$ arcsec), which is younger and more metal-rich than what we find. \citet{chilingarian09} found an age of $3.8$ to $6.7$ Gyr with $[Z/Z_{\odot}]=-0.55$ or $-0.82$, within $11.4$ arcsec; this is consistent with our results. \citet{paudel10} found that the dE is $5.2^{+1.7}_{-1.9}$ Gyr old with $[Z/Z_{\odot}]=-0.61\pm0.25$ within $19.1$ arcsec, which is also marginally consistent with our results. \citet{michielsen08} derived a few age and metallicity estimates based on different stellar population models, and generally found old ages and intermediate sub-solar metallicities, consistent with our results. 

\subsection{UDG DGSAT I}\label{sec:DGSAT I}
For the UDG DGSAT I, we similarly carry out \texttt{prospector} SED fitting in two scenarios: (1) assume a uniform prior over $A_V=0-4$ mag, and (2) fix $A_V=0$ mag. The second scenario is to mimic what is common practice in the literature, namely to ignore dust altogether (we do not have IRAC4 data as for the dE above). The highest likelihood model spectra in the two scenarios are shown in \autoref{fig:seds}. Again, the best-fit models in both scenarios agree with the optical--NIR data but diverge dramatically at longer wavelengths. Interestingly, even with $A_V=0$ mag, the highest likelihood model spectrum predicts mid-IR dust emission features, presumably from TP-AGB and/or post-AGB stars. 

The marginalized posteriors for the free parameters are shown in \autoref{fig:posteriors}. Regardless of our dust treatment, this UDG is consistent with a systematically younger age compared to the dE, with its asymmetric posterior peaking at $\sim3$ Gyr and exhibiting a long tail toward older ages (but a sharp cut-off at very young ages). The metallicity posterior in the scenario with dust left as a free parameter peaks at sub-solar metallicities ($[Z/Z_{\odot}]\approx-0.5$). In contrast, when we assume exactly zero diffuse interstellar dust, the metallicity posterior is still consistent mostly with sub-solar metallicities but is systematically shifted toward higher values (as expected), with a tail toward solar and super-solar metallicities. We give summary statistics for the free parameters in the two scenarios in \autoref{tab:freeparams} and \autoref{tab:freeparamsnodust}.

We note that our total stellar mass (provided in \autoref{sec:mstartot}) is only $\sim1.2\times$ higher than the total stellar mass reported by \citet{martinezdelgado16}. This slight difference is likely due to a combination of three things: (1) we have different bad pixel masks, (2) our stellar mass is based on fully Bayesian optical--NIR SED fitting, and (3) simple color-dependent stellar $M/L$ ratios, such as that used by \citet{martinezdelgado16}, often have a factor of two uncertainty anyway \citep[e.g.,][]{bell01}. For reference, \citet{martinezdelgado16} estimated $M_*/L_I=1.1M_{\odot}/L_{\odot}$ based on their GALFIT model to the entire UDG, whereas within our aperture of radius 15 arcsec (roughly their $R_e$) we find $M_*/L_I=1.3$ (using the median stellar mass in the scenario with dust, which is very similar to the case without dust). 

\subsection{UDG VCC 1287}\label{sec:VCC 1287}
\begin{figure*} 
\begin{center}
\includegraphics[width=\hsize]{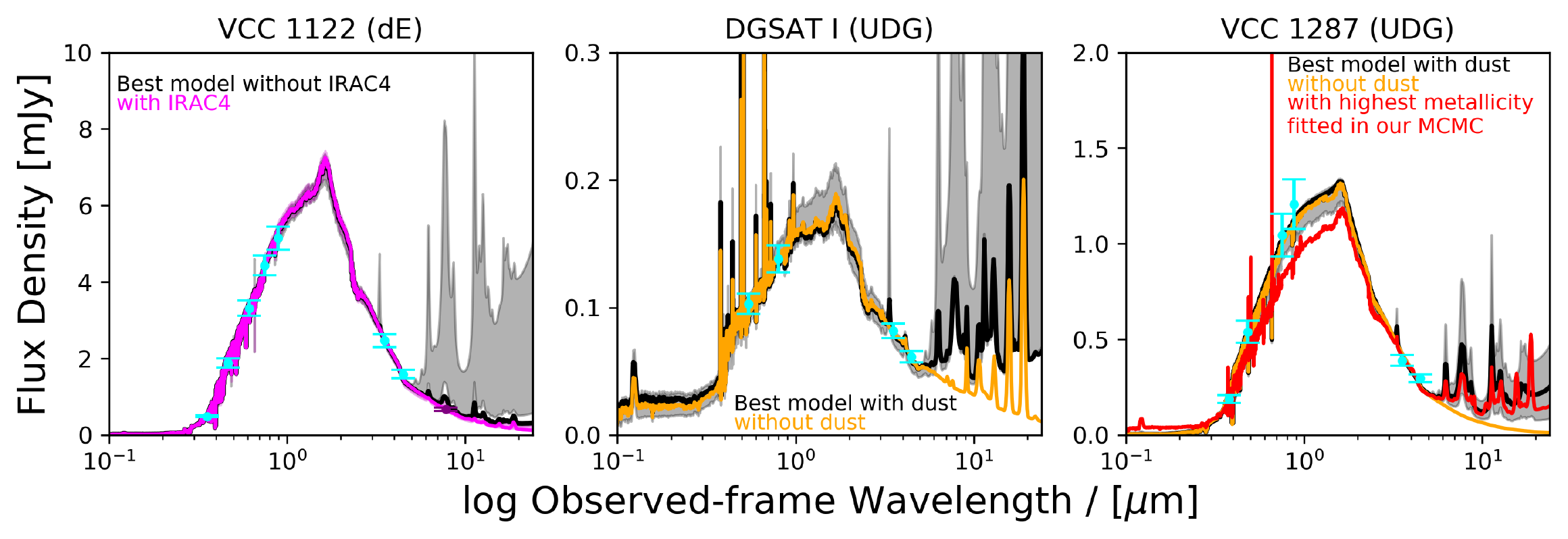}
\end{center}
\caption{Observed photometry and \texttt{prospector} model spectra for the dE VCC 1122 (left),  UDG DGSAT I (middle), and UDG VCC 1287 (right). The wavelength range shown is $0.1-24\mu$m. The cyan points show our observed photometric measurements, with the IRAC4 data point for the dE shown in purple. The thick black line shows the highest likelihood spectrum when diffuse interstellar dust is left as a free parameter with uniform prior (and in the case of the dE, when the IRAC4 data point is excluded from the fit). The gray shading reflects the 16-84th percentile spread in flux density at each wavelength in that scenario with dust. For the dE, the magenta line shows the best model when IRAC4 is included in the fit and dust is allowed \textbf{(magenta shading also shows the 16-84th percentile spread in these model predictions, which are much tighter than without IRAC4 and no longer extend to high MIR flux densities)}. For the UDGs, the orange line shows the highest likelihood spectrum when we instead assume zero diffuse interstellar dust. For VCC 1287, we also show in red the spectrum with the highest fitted metallicity $[Z/Z_{\odot}]=-0.9$ from our MCMC (with dust allowed); clearly it is not a good fit to the reddest optical data. Note how for DGSAT I, even with diffuse interstellar dust turned off, the SPS models predict strong mid-IR emission features (presumably from TP-AGB and/or post-AGB stars).}
\label{fig:seds}
\end{figure*}

\begin{figure*} 
\begin{center}
\includegraphics[width=\hsize]{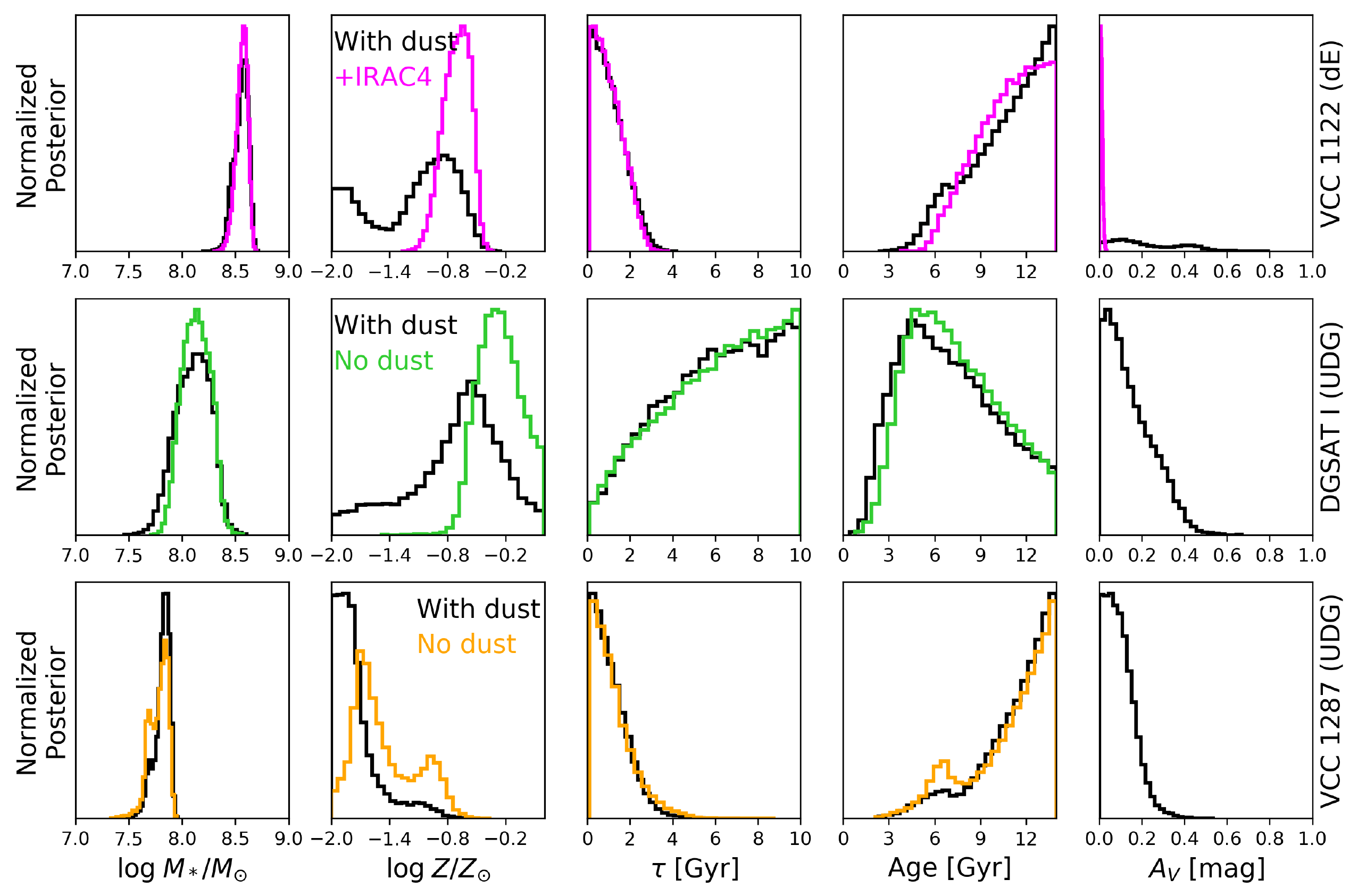}
\end{center}
\caption{The marginalized posteriors for each of the five free parameters (all with uniform priors) in two different scenarios for how we treat diffuse interstellar dust. In black, the marginalized posteriors are for the case where $A_V$ is left as a free parameter with uniform prior over $0-4$ mag. The marginalized posteriors for the case where $A_V$ is fixed to 0 mag (i.e., no diffuse interstellar dust) are shown in green (DGSAT I) and orange (VCC 1287). Instead of fixing $A_V=0$ mag for the dE, we include IRAC4 in the fit and retain a uniform prior of $A_V=0-4$ mag (the magenta posteriors). The stellar masses are aperture masses. For this particular dE, IRAC4 is effective in ruling out models with high dust optical depths. The strong dust--metallicity degeneracy is quite apparent: with exactly zero dust, a higher metallicity is possible (though both UDGs are still consistent with sub-solar metallicities). The full covariances between free parameters are shown in Appendices \ref{sec:covdust} and \ref{sec:covnodust} for the two dust scenarios.}
\label{fig:posteriors}
\end{figure*}

Our SED fitting for the UDG VCC 1287 was done with the same two scenarios as for DGSAT I: (1) let $A_V$ be a free parameter with uniform prior over $0-4$ mag, and (2) fix $A_V=0$ mag. In \autoref{fig:seds}, we show the highest likelihood spectra for the two dust scenarios. For the first scenario with dust, we also separately show the model spectrum with the highest predicted stellar metallicity ($[Z/Z_{\odot}]\approx-0.8$). The MCMC model with the highest stellar metallicity clearly does not fit the reddest optical data well, and demonstrates that this UDG is exceptionally metal-poor. Indeed, the metal-poor nature of this UDG is strikingly suggested by the marginalized posteriors for stellar metallicity shown in \autoref{fig:posteriors}. Even when assuming zero diffuse interstellar dust, the metallicity posterior becomes bimodal and shifts toward higher values but still is more likely to have $[Z/Z_{\odot}]\approx-1.7$ than $[Z/Z_{\odot}]\approx-1.0$. In both dust scenarios, VCC 1287 appears to be old ($>7.5$ Gyr) and metal-poor (see \autoref{tab:freeparams} and \autoref{tab:freeparamsnodust}).

Both our aperture stellar mass and our total stellar mass are significantly higher than the values reported in \citet{beasley16}. We defer an explanation of this to Appendix \ref{sec:elixir} and Appendix \ref{sec:galfit}, but briefly mention here that VCC 1287 was subject to background over-subtraction in the public archival CFHT imaging used by \citet{beasley16}. Using the dynamical masses reported by \citet{beasley16}, here we provide revised estimates of the dynamical to stellar mass ratio within their $R_e$ and in total. \citet{beasley16} measured $M_{\rm dyn}(<R_e)\sim2.6\times10^9M_{\odot}$ and total $M_{\rm dyn}\sim8\times10^{10}M_{\odot}$. Thus, we now find $M_{\rm dyn}/M_*\approx41$ within their $R_e=30$ arcsec (a factor of $\sim2$ lower than their value of $\sim93$), and $M_{\rm dyn}/M_*\approx400$ using total masses (a factor of $\sim7.5$ lower than their value of $\sim3000$). While these values are less extreme than those reported in \citet{beasley16}, our revised estimates still suggest that VCC 1287 has an elevated dynamical mass for its stellar mass and luminosity.

\begin{table}
\caption{\texttt{prospector} Results With Dust}
\centering
\begin{tabular}{cccc}
\hline
Parameter & VCC 1122 & DGSAT I & VCC 1287 \\\hline
$\log M_*/M_{\odot}$ & $8.56^{+0.06}_{-0.08}$ & $8.12^{+0.16}_{-0.18}$ & $7.82^{+0.05}_{-0.10}$ \\
$[Z/Z_{\odot}]$ & $-1.08^{+0.35}_{-0.73}$ & $-0.63^{+0.35}_{-0.62}$ & $<-1.55$ \\
$\tau$ [Gyr] & $<1.83$ & $>3.20$ & $<1.93$ \\
$t_{\rm age}$ [Gyr] & $>7.86$ & $6.81^{+4.08}_{-3.02}$ & $>8.66$ \\
$A_V$ [mag] & $0.18^{+0.24}_{-0.12}$ & $<0.26$ & $<0.16$ \\
\end{tabular}
\tablecomments{Summary statistics for the marginalized posterior of each free parameter shown in \autoref{fig:posteriors}, in the case where diffuse interstellar dust is fit as a free parameter with a uniform prior of $A_V=0-4$ mag. In cases where the posterior distribution is skewed and hitting up against the prior limit (e.g., $t_{\rm age}=14$ Gyr), we instead give either the lower or upper limit (16th or 84th percentile, respectively). Otherwise, the 16th, 50th and 84th percentiles of the posterior distribution are given. Note that the stellar masses given here are aperture masses; total stellar masses derived via GALFIT are given in \autoref{sec:mstartot}.}
\label{tab:freeparams}
\end{table}

\begin{table}
\caption{\texttt{prospector} Results With Minimal/No Dust}
\centering
\begin{tabular}{cccc}
\hline
Parameter & VCC 1122 & DGSAT I & VCC 1287 \\\hline
$\log M_*/M_{\odot}$ & $8.57^{+0.05}_{-0.06}$ & $8.13^{+0.14}_{-0.14}$ & $7.80^{+0.07}_{-0.11}$ \\
$[Z/Z_{\odot}]$ & $-0.70^{+0.14}_{-0.16}$ & $-0.27^{+0.25}_{-0.22}$ & $-1.56^{+0.52}_{-0.19}$ \\
$\tau$ [Gyr] & $<1.76$ & $>3.21$ & $<2.12$ \\
$t_{\rm age}$ [Gyr] & $>8.64$ & $7.12^{+3.79}_{-2.79}$ & $>7.74$ \\
$A_V$ [mag] & $<0.02$ & $-$ & $-$ \\
\end{tabular}
\tablecomments{Same as \autoref{tab:freeparams} but now we assume zero diffuse interstellar dust ($A_V$ fixed to 0 mag) for the two UDGs. For the dE, instead of fixing $A_V=0$ mag, we include IRAC4 in the fit and assume a uniform prior over $A_V=0-4$ mag. The vastly reduced confidence interval for $A_V$ \textbf{(indeed, it is now an upper limit)} shows that IRAC4 alone is quite helpful for ruling out high dust optical depths within this particular dE.}
\label{tab:freeparamsnodust}
\end{table}

\section{Discussion}\label{sec:discussion}
\subsection{The Stellar Populations of Our UDGs}\label{sec:stellarpops}
The stellar populations of our two UDGs appear to be quite different relative to each other and compared to our comparison dE. This was evident already when comparing all three objects to single SSP evolutionary tracks in optical--NIR color--color diagrams (see \autoref{fig:colorcolor}). It also appears to be the case after marginalizing over complicated degeneracies in the stellar population models, particularly dust and extended SFHs, using \texttt{prospector}. Taking into account the IRAC4 data, the Virgo dE is consistent with a relatively old stellar population with intermediate sub-solar metallicity (and relatively little diffuse interstellar dust). The Virgo UDG also appears to have a similarly old age and small $\tau$ (i.e., a similar SFH that could be approximated by an SSP), but its stellar metallicity may be even lower than that of the dE. If we assume exactly zero diffuse interstellar dust in the Virgo UDG, then its metallicity may be consistent with that of the Virgo dE. On the other hand, while the metallicity of DGSAT I appears consistent with that of the dE (and perhaps the Virgo UDG), the marginalized posteriors for its age and $\tau$ are not consistent with those of either the Virgo dE or the Virgo UDG. Furthermore, if we assume no diffuse interstellar dust at all in DGSAT I, then its stellar metallicity posterior is even consistent with solar values. 

The above results are interesting in light of the fact that DGSAT I lives in a low density environment \citep{martinezdelgado16} unlike VCC 1287 and VCC 1122. Under the assumption of an exponentially decaying SFH, the number of e-folds ($t_{\rm age}/\tau$) is a proxy for the ratio of young to old stars, or the mean stellar age. Both Virgo objects have undergone a similarly large number of e-folds, with the dE having $10.5^{+18.4}_{-3.2}$ and the UDG VCC 1287 having $9.7^{+16.2}_{-5.0}$. In contrast, the field UDG DGSAT I is consistent with only $1.2^{+1.0}_{-0.4}$ e-folds. Furthermore, the two Virgo objects might more or less be consistent with single burst SSP scenarios since $\tau$ is an upper limit consistent with very small values, but the field UDG probably had a more complicated (potentially bursty) SFH since its $\tau$ posterior is not asymmetrically peaked toward very low values. In fact, the covariance between $t_{\rm age}$ and $\tau$ for DGSAT I seen in \autoref{fig:cornerDGSAT I} and \autoref{fig:cornerDGSAT Ib} (depending on dust scenario) suggests that the older it is, the more extended its SFH was (i.e., a burst that occurred long ago would keep forming stars until relatively recently, whereas a burst that happened relatively recently would be truncated more rapidly). 

How do the stellar populations of our UDGs compare to those in the literature? Direct observational constraints for the ages and metallicities of red UDGs are sparse, especially those obtained via explicit fitting and marginalization of some sort.\footnote{There has been some spectroscopic work on a few ``blue UDGs" that have HII regions indicative of very young stellar populations, enabling gas-phase metallicity constraints \citep[generally sub-solar gas-phase metallicity;][]{trujillo17,bellazzini17}. However, we are only considering red UDGs in this paper.} \citet{vandokkum15a} initially pointed out the age--metallicity degeneracy for their red Coma UDGs, and preferred old and sub-solar metallicity stellar populations on the basis of the median $g-i$ color of their sample (ignoring dust). \citet{roman17b} used two optical colors ($g-r$ and $g-i$) to more fully map out the range of allowed age and metallicity combinations but could not rule out young metal-rich populations \citep[again, also ignoring dust; see also][]{vanderburg16}. Interestingly, the $g-i=0.72$ AB mag color of VCC 1287 is similar to the median $g-i=0.8$ color of Coma UDGs reported by \citet{vandokkum15a} and to the average $g-i=0.75$ AB mag color found by \citet{roman17b} for their red UDGs. As for DGSAT I, \citet{martinezdelgado16} showed that a single SSP with intermediate age ($1.7\pm0.4$ Gyr) and roughly solar metallicity ([Fe/H]$=-0.2\pm0.3$) was a good fit to their long slit spectrum; our SED fitting is marginally consistent with their interpretation (even after taking into account dust and SFHs). However, we caution that the exact placement of their long slit, the contribution of foreground/background contaminants, and the contribution of the DGSAT I ``overdensity" in particular, are unclear. 

In terms of spectroscopy, \citet{kadowaki17} detected Balmer and Ca H\&K absorption lines using deep spectroscopy for four Coma cluster UDGs. By visually comparing their stacked spectrum to four single SSPs (1 Gyr old with [Fe/H]$=0.2$, and 13 Gyr old with [Fe/H]$=-0.5$, $-1.5$ and $-2.0$), they concluded that their four UDGs were broadly consistent with an old metal-poor ([Fe/H]$\lesssim-1.5$) stellar population. Recently, \citet{gu17} carried out full (optical) spectral fitting using ultra-deep integral field unit (IFU) spectroscopy of three Coma UDGs and concluded that their UDGs were consistent with low metallicities and old ages (allowing for a single burst scenario, but neglecting dust). Specifically, \citet{gu17} found average ages of $\sim8-9$ Gyr and average [Fe/H] from $-1.3$ to $-0.8$. Our results for the Virgo UDG VCC 1287 are consistent with those values (especially in the case of zero dust). On the other hand, DGSAT I has an age posterior that peaks at a systematically younger age ($\sim3$ Gyr) and a metallicity posterior that likewise is consistent with higher values (especially in the case of no dust) than the Coma UDGs studied by \citet{gu17}. 

\begin{figure*} 
\begin{center}
\includegraphics[width=0.6\hsize]{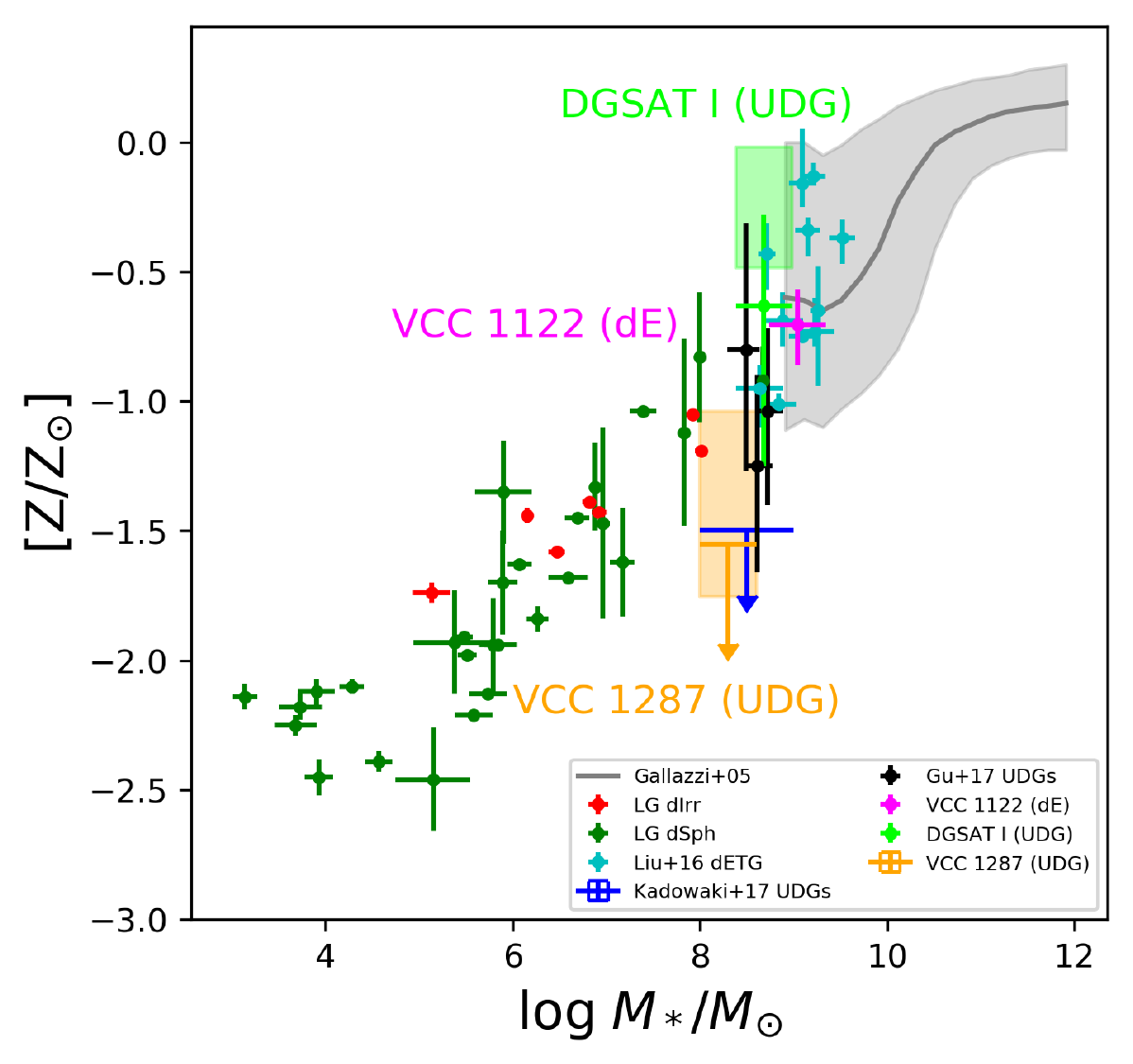}
\end{center}
\caption{The stellar mass--metallicity relation. For comparison, we plot Local Group dIrrs (red points) and dSphs (green points); see \citet{kirby13}. Virgo dwarf early-type galaxies from \citet{liu16} are shown in cyan. The mass--metallicity relation extending to more massive galaxies (both star-forming and quiescent) from \citet{gallazzi05} is plotted in grey. The spectroscopic results on Coma UDGs from \citet{kadowaki17} and \citet{gu17} are shown in blue and black, respectively. The magenta point shows our results for the dE VCC 1122 with IRAC4 included in the fit. The green and yellow points show the results for DGSAT I and VCC 1287, respectively, in the scenario with dust allowed. The shaded green and yellow regions show how the results for the UDGs would change if $A_V=0$ mag exactly. In the scenario with dust, the stellar metallicity measurement for VCC 1287 is considered an upper limit. We use total GALFIT-based stellar masses for our objects in this plot (see \autoref{sec:mstartot}).}
\label{fig:mzr}
\end{figure*}

While both of our UDGs are broadly consistent with sub-solar metallicities, their marginalized posterior distributions for $[Z/Z_{\odot}]$ are quite different from each other (and from the dE VCC 1122, regardless of whether IRAC4 is included in the fit; see again \autoref{fig:posteriors}). We can effectively rule out solar and super-solar stellar populations for the Virgo UDG, but not for the field UDG DGSAT I (which depends on the diffuse interstellar dust content that is poorly constrained by the available data). If DGSAT I has a metal-rich stellar population, that would be surprising given the expectation from the stellar mass--metallicity relation (MZR) for dwarfs. In \autoref{fig:mzr}, we overplot the two UDGs and the dE on the stellar MZR using their total GALFIT-based stellar masses (see \autoref{sec:mstartot}). For comparison, we include measurements of stellar metallicity and total stellar mass for Local Group dwarfs of different types (MW dSphs, M31 dSphs, and Local Group dIrrs) from \citet{kirby13}. We also include the sample of more massive Virgo dETGs from \citet{liu16}, and the \citet{gallazzi05} trend for massive galaxies more generally. Finally, we also include the Coma UDG results from \citet{kadowaki17} and \citet{gu17}. Note that \texttt{prospector} assumes solar abundances (i.e., [$\alpha$/Fe]$=0$ and $[Z/X]=$[Fe/H]), so where necessary we assume that literature [Fe/H] values are simply $[Z/Z_{\odot}]$.

The degree to which our UDGs are consistent with the stellar MZR for dwarfs depends on our treatment of dust. If the normalization of the dust attenuation curve is allowed to be a free parameter with uniform prior over $A_V=0-4$ mag, then DGSAT I is consistent with the MZR but VCC 1287 is exceptionally metal-poor. On the other hand, if we fix $A_V=0$ mag exactly, then VCC 1287 is more or less consistent with the MZR but DGSAT I is on the high end. In either dust scenario, the Virgo cluster UDG VCC 1287 has a low stellar metallicity, which is in agreement with the Coma UDG results of both \citet{kadowaki17} and \citet{gu17}. The higher inferred stellar metallicity of DGSAT I may be related to its location within a low density environment and its extended SFH \citep[and correspondingly higher stellar feedback; see][]{dicintio17}. 

\subsection{Implications for UDG Formation}\label{sec:formation}
A key unanswered question about UDGs is whether they are simply a continuous extension of the normal dwarf population \citep[][]{rong17,gu17}. A ``differential comparison" between the stellar populations (i.e., ages and metallicities) of UDGs and typical dwarfs can help answer this question. Although our sample size is small, here we comment on what our results suggest about the formation of UDGs by qualitatively comparing to existing theoretical predictions. 

In the scenario where UDGs purely represent the high spin tail of the typical dwarf population \citep{amorisco16}, there are no specific predictions for ages and metallicities but we can assume that the stellar population properties of UDGs are continuous with those of typical dwarfs. In other words, UDGs should follow the stellar MZR and any other correlations between SFH (and hence age) with environment. We discussed in \autoref{sec:stellarpops} that DGSAT I may be consistent with the stellar MZR (albeit a bit high if there is no diffuse interstellar dust) and that it also shows evidence of a more extended SFH. Thus, DGSAT I may be compatible with the high spin dwarf scenario (assuming it is a face-on disk). Concrete evidence to the contrary would come from constraints on its halo mass and halo spin parameter. VCC 1287 is more complicated because if it contains even marginal amounts of diffuse interstellar dust, then its stellar metallicity is likely exceptionally low. Otherwise, VCC 1287 is more or less consistent with the expectation for dwarfs, though we note that it has an elevated dynamical to stellar mass ratio \citep[][and see our \autoref{sec:VCC 1287}]{beasley16}.

Building on the exclusively high spin scenario, \citet{rong17} made quantitative predictions for the mass-weighted ages of UDGs compared to typical dwarfs (dEs and dSphs), taking into account different environments as well as the infall times of UDGs and dwarfs into clusters. Their study was based on the \citet{guo11} semi-analytic model applied to dark matter halo merger trees extracted from the MSII \citep{boylankolchin09} and Phoenix \citep{gao12} N-body simulations, which together probe a diverse range of environments (from field environments to massive Coma cluster analogs). According to their model, red UDGs follow a hybrid formation scenario of high spin host halos as well as ``late formation" times. It is important to clarify that ``late formation" can mean two separate physical processes. First, UDGs may have fallen into clusters at later times compared to normal dwarfs, and thus their star formation was quenched more recently (leading to relatively younger ages for UDGs). Second, if UDGs kept undergoing merger-induced star formation at lower redshifts compared to typical dwarfs, then a natural consequence of the overall decreasing cosmic mean density is that merger remnants would be more diffuse at low redshift compared to high redshift \citep[see also][and references therein]{porter14}. Under the hybrid scenario of \citet{rong17}, UDGs should be systematically younger than normal dwarfs by $\sim2.5$ Gyr. 

Given our limited data and the complicated age--metallicity--dust degeneracy, we cannot robustly determine whether VCC 1287 is significantly younger or older than VCC 1122.\footnote{Without UV, MIR/FIR and emission line data, we can only constrain ages down to the $\sim1$ Gyr level. We cannot constrain very recent star formation (on the order of 10 Myr or 100 Myr). Furthermore, with our optical--NIR broad-band photometry alone, we also cannot precisely pin down very old ages such as distinguishing between an 8 Gyr old and a 12 Gyr old stellar population.} What we can say is that both objects are predominantly old and that their best-fit SFHs are not significantly extended. Thus, it is unlikely that VCC 1287 is consistent with the ``late formation" aspect of the \citet{rong17} hybrid scenario. In contrast, the extended SFH and relatively younger age of DGSAT I compared to both of the Virgo objects suggests that it might be consistent with at least the ``late formation" part of the \citet{rong17} scenario (whether it has a high halo spin is currently unknown). We caution that a larger observational sample of both UDGs and typical dwarfs is required to comment on the \citet{rong17} scenario in a proper statistical and cosmological context. Specifically, a detailed stellar population analysis such as ours and that of \citet{gu17} of both blue and red UDGs, using a control sample of normal dwarfs with similar luminosities and colors, as a function of clustercentric distance, might be a fruitful next step \citep[see also][]{roman17a,vanderburg17}. 

While our data are inadequate for exploring the stellar population properties of the DGSAT I ``overdensity" pointed out by \citet{martinezdelgado16} (see our \autoref{sec:maskbkg}), that might be additional evidence for an extended and potentially bursty SFH. This would support the ``late formation" aspect of the \citet{rong17} scenario. Furthermore, \citet{dicintio17} found an analog of DGSAT I in the NIHAO hydrodynamical simulations in terms of size and surface brightness, and showed that the ``overdensity" could result from a second, younger stellar population component. In their simulations, UDGs that live in low density field environments are exceptionally large due to stellar feedback-driven expansion and radial stellar migration. While they did not predict stellar metallicities, it is reasonable to expect that an elevated amount of stellar feedback would also enrich the interstellar medium with more metals, leading to a higher gas and stellar metallicity \citep[see also][]{somerville15}. Under the \citet{dicintio17} scenario, DGSAT I could be expected to lie above the stellar MZR, which is what we see in \autoref{fig:mzr} (especially in the case with zero dust). Future IFU spectroscopic constraints on the gas and stellar metallicities, and non-parametric SFH, of DGSAT I would help further test the \citet{dicintio17} expanded dwarf scenario.

We now turn to the intriguing possibility that red UDGs might be ``failed" galaxies, a scenario first suggested by \citet{vandokkum15a} and theoretically explored by \citet{yozin15}. This scenario requires that UDGs have overmassive dark matter halos given their dwarf-like stellar masses, which might be plausible for VCC 1287 \citep{beasley16} but is still unconstrained for DGSAT I. Specifically, the ``failed" galaxy scenario, based on the theoretical work of \citet{yozin15}, would predict very old mean stellar ages of $\sim10$ Gyr (assuming the UDG progenitor began its infall toward a cluster at $z=2$ and had a quenching timescale of $\tau\approx2$ Gyr, due to gas stripping). Sub-solar metallicities around $[Z/Z_{\odot}]\approx-1.5$ were assumed for this scenario following \citet{vandokkum15a} because it is likely that metal production was suppressed due to early quenching. VCC 1287 appears consistent with the failed galaxy picture, similar to the red Coma UDGs analyzed by \citet{gu17} and \citet{kadowaki17}. We can likely rule out the ``failed" galaxy scenario for DGSAT I based on its relatively younger age and its relatively high metallicity (potentially solar if there is very little diffuse interstellar dust). 

The potentially very low metallicity of VCC 1287 would seemingly invite more exotic scenarios for its formation. One tantalizing possibility is that VCC 1287 could be an analog of DF17 \citep{peng16,beasley16b} and many Coma UDGs \citep{vandokkum17}, which were found to host a large number of GCs \citep[photometrically-selected GC candidates in the case of DF17 and Coma UDGs, but spectroscopically confirmed GCs for VCC 1287;][]{beasley16}. It is interesting to wonder whether the low metallicity of VCC 1287 implies a direct link to its GC system, and whether VCC 1287 might also be some sort of pure stellar halo as originally suggested for DF17 \citep[see also][for a related discussion about the Fornax dSph]{larsen12}. \citet{peng16} qualitatively discussed a starburst scenario in which GC progenitors form first, use up and drive away most of the leftover gas, and quickly halt any subsequent formation of new field stars \citep[and in the process, prevent the build-up of any significant disk or bulge structural component; see also][]{katz13}. This scenario would manifest in a rapidly truncated SFH and $\alpha$-element enhancement for the field star population. Conditional on our assumption of an exponential SFH model, the e-folding timescale $\tau$ is relatively short for VCC 1287 and thus consistent with this scenario. Modeling $[\alpha/$Fe$]>0$ is beyond the scope of our SED modeling efforts. 

Finally, when it comes to phenomena involving tidal forces, we consider two distinct possibilities: (1) tidal debris formed as a byproduct of galaxy mergers, and (2) tidally disrupted normal dwarfs. Tidal debris is generally thought to be very blue and young because the vast majority of such tidal dwarfs are found near gas-rich or mixed merger remnants \citep[typically only $\sim100$ Myr old; see][and references therein]{kaviraj12}. However, tidal debris can still be relatively old and metal-rich if it is torn off of existing massive galaxies in high density environments, or if it is ejected during gas-poor mergers \citep[e.g.,][]{vandokkum05}. \citet{greco17a} proposed that their ``Sumo Puff" UDG candidate is consistent with tidal debris, owing to the tentative discovery of a bridge of material connecting their UDG candidate to a nearby (on-sky) post-merger galaxy. In both tidal scenarios that we consider, the mean stellar ages and metallicities of the tidal material would be expected to follow those of their progenitors (assuming negligible star formation). We can probably rule out a direct tidal origin altogether for the field UDG DGSAT I. VCC 1287 is more complicated because the spectroscopic redshift of its nucleus suggests it is within the Virgo cluster. \citet{beasley16} disfavored the idea that VCC 1287 is tidal debris based on its exceptionally high dark matter halo mass. Since tidal debris would be expected to lie above the stellar MZR in a cluster environment, our low derived stellar metallicity also argues against the tidal debris scenario for VCC 1287. As for whether VCC 1287 is a tidally disrupted normal dwarf, an exceptionally low stellar metallicity would be hard to reconcile with the generally intermediate sub-solar metallicities of dEs \citep[e.g.,][]{liu16}. VCC 1287 also has an overabundant population of GCs compared to normal dwarfs \citep{beasley16}. In the future, it might be fruitful to compare the stellar populations of red UDGs and dSphs, derived in a similar and self-consistent way \citep[see also][]{mcconnachie12,collins13,makarov15,toloba16,crnojevic16,merritt16}.

\subsection{Systematic Uncertainties and Future Prospects}\label{sec:systematics}
In \texttt{prospector}, we made several prior assumptions about the physical properties of our UDGs such as the shape of their internal dust attenuation curve, a parametric form for their SFH, and linearly uniform priors on all free parameters. In an effort to assess the impact of these assumptions on our results, we re-ran \texttt{prospector} several times for both UDGs, each time changing a single assumption (and always working in the scenario with dust allowed). The things we varied were: (1) use an SMC-like dust attenuation curve since it has a steeper power law slope than the one in \citet{calzetti00}, (2) use a MW dust attenuation curve with $R_V=A_V/E(B-V)=3.1$ fixed to the MW value, since this curve includes the 2175\AA\ Si absorption feature, (3) use a logarithmic uniform prior over $\tau=0.1-100$ Gyr instead of a linear uniform prior over $\tau=0.1-10$ Gyr to account for how quickly the SED changes shape as a function of $\tau$, and (4) use a delayed-$\tau$ SFH model instead of the normal $\tau$ model, with a logarithmic prior over $\tau=0.1-100$ Gyr. In all of the tests that we ran, the parameter covariances and marginalized posteriors were very similar to our baseline results. While this suggests that our results are insensitive to the adopted prior assumptions, we caution that much more data are needed on UDGs in order to derive their physical properties in a model-independent and non-parametric way.

Furthermore, we enabled nebular emission lines by default in \texttt{prospector} according to the prescription of \citet{byler17}. Some of the MCMC spectra in \autoref{fig:seds} have optical emission lines. These optical emission lines come from star formation (optical emission lines from post-AGB stars in old stellar systems are not implemented in \texttt{prospector} for computational efficiency reasons and because their contribution is expected to be small). At least in the case of DGSAT I, the emission lines from young stars are consistent with its inferred age and e-folding time (i.e., $t_{\rm age}/\tau\sim1$). The nebular emission lines are a smaller source of systematic error on derived physical properties such as $[Z/Z_{\odot}]$ compared to other assumptions, particularly AGB dust models and solar chemical abundance patterns. For example, if $[\alpha/$Fe$]\neq0$, that would reasonably translate to a systematic uncertainty of at most $\sim0.3$ dex on $[Z/Z_{\odot}]$.\footnote{However, many of the relevant metal absorption lines are at bluer wavelengths (namely in the UV), and therefore redder optical and NIR broadband photometry would be insensitive to abundance variations. Indeed, using mock spectra from the \texttt{alf} code \citep[see][]{conroy18} for a $10$ Gyr old stellar population with $[Z/Z_{\odot}]=-1.5$, we find negligible changes in broadband magnitudes for the CFHT $u^*giz$ and $JHK$ NIR bandpasses (covering the wavelength range of the mock spectra) when $[\alpha/$Fe$]=0$ and $[\alpha/$Fe$]=0.3$ \citep[a typical enhanced value for dwarfs; e.g.,][]{liu16}.} While this would not be enough to make VCC 1287 consistent with solar or super-solar metallicities, it is a potentially significant source of systematic error for DGSAT I. In the future, ultra-deep spectroscopy can provide constraints on both of these fronts, with information about the existence of warm ionized gas in these objects as well as any possible chemical abundance variations. 

We emphasize that our results are highly dependent on using the combined optical--NIR SEDs of our objects rather than the optical SEDs alone. This is because we cannot break the age--metallicity degeneracy without the $Spitzer$-IRAC data (dust further complicates this problem). To see how strongly our results depend on the NIR data, we re-ran \texttt{prospector} with our fiducial code setup (\autoref{sec:prospector}) for the Virgo UDG VCC 1287 using only its CFHT $u^*giz$ optical SED (a similar test for DGSAT I is not as useful because we only have two optical bandpasses). In both of our dust scenarios, the aperture stellar mass is similar to within $\sim0.1$ dex compared to the original run with IRAC data included. However, the metallicity posterior is much broader and extends to solar and super-solar values, even in the case where we assume $A_V=0$ mag exactly. The marginalized posteriors for age and $\tau$ are also significantly different, with significantly younger ages ($\sim4.5$ Gyr) and larger e-folding times ($\sim7.5$ Gyr) allowed in the case with dust. Finally, the $A_V$ posterior itself also extends to values greater than 1 mag, which was effectively ruled out with the IRAC NIR data. All of this confirms that the NIR data is playing a crucial role in our analysis. Since NIR emission is strongly affected by models for AGB circumstellar dust emission \citep[e.g.,][]{villaume15} as well as NIR stellar absorption features \citep[e.g.,][]{peletier12,norris14}, our results could be affected by systematics related to modeling broad-band NIR emission \citep[e.g.,][and references therein]{villaume17}. In a similar vein, we have not taken into account variations in the morphology of the blue horizontal branch, which for old, metal-poor systems could result in artificially younger light-weighted mean ages \citep[][and references therein]{conroy18}.

Another limitation of our study, and hence an avenue for future work, is that we are restricted in the range of $[Z/Z_{\odot}]$ that we can explore because we are using the Padova isochrones \citep{marigo07,marigo08}, which only span $-2.0<[Z/Z_{\odot}]<0.2$. Thus, we cannot test whether, e.g., VCC 1287 is consistent with even lower metallicity values as is the case for many nearby low-mass dwarfs (see again \autoref{fig:mzr}). In the future, it will be particularly interesting to explore the case of extremely low metallicities, especially in the case of VCC 1287. Constraining the normalization and shape of the SED peak with $JHK$ photometry might help pin down the metallicity better. We already found that the observed photometry of VCC 1287 is not described well by the highest metallicity MCMC spectrum (with $[Z/Z_{\odot}]=-0.9$; magenta line in \autoref{fig:seds}) because it predicts less flux at the reddest optical wavelengths than is observed. Assuming the reverse trend occurs for spectra with $[Z/Z_{\odot}]<-2.0$, very metal-poor models might produce a steep increase in flux toward the NIR SED peak that $JHK$ photometry could map out.

Finally, we caution that effectively nothing is directly known about the circumstellar and diffuse interstellar dust content of UDGs. In this paper, we have taken the first steps to explicitly characterize the normalization of several different assumed attenuation curves, and found that both UDGs are relatively dust-free with $A_V\lesssim0.5$ mag. However, our results are based on optical--NIR SED fitting, and it is well known that the optical--NIR SED shape itself does not provide robust constraints on the diffuse interstellar dust. Photometry in the UV, MIR and ideally FIR (near the $160\mu$m diffuse dust emission SED peak), or spectroscopic constraints via the Balmer decrement or MIR emission lines, would be more suitable. While these other constraints are likely too expensive to obtain for red UDGs such as the two objects considered in this paper, they could offer valuable information about diffuse interstellar dust within so-called ``blue UDGs." Nevertheless, marginalizing over dust attenuation is crucial for tackling the age--metallicity--dust degeneracy in red UDGs; e.g., for all three of our objects, the MCMC ``corner" plots in Appendix \ref{sec:covdust} reveal that even with $A_V\lesssim0.5$, the dust--metallicity degeneracy is strong and extends over the full range of stellar metallicity probed. Future instruments such as the Mid-Infrared Instrument (MIRI) aboard the \textit{James Webb Space Telescope} may allow us to place stronger priors on the diffuse interstellar dust content of the nearest representative red UDGs. 

\section{Summary}\label{sec:summary}
We have presented Bayesian optical--NIR SED fitting results for two UDGs and one normal cluster dE. The UDGs are both optically red, but live in quite different environments: DGSAT I is in the field ($\sim2$ Mpc in projection away from a cluster) and VCC 1287 is in the Virgo cluster. Our results can be summarized as follows: 

\begin{enumerate}
\item The Virgo UDG VCC 1287 is consistent with an old ($\gtrsim7.7$ Gyr) and metal-poor ($[Z/Z_{\odot}]\lesssim-1.0$) stellar population after marginalizing over diffuse interstellar dust uncertainties. When assuming an exponentially-declining SFH model, we cannot rule out an effectively single SSP for VCC 1287, suggesting that it is indeed very old.
\item DGSAT I appears to be systematically younger than the Virgo UDG, with an age posterior extending down to $\sim3$ Gyr. After marginalizing over uncertainties in diffuse interstellar dust content, DGSAT I appears to have a higher stellar metallicity than the Virgo UDG, with $[Z/Z_{\odot}]=-0.63^{+0.35}_{-0.62}$. If we assume exactly zero diffuse interstellar dust, DGSAT I might even be consistent with a solar metallicity stellar population (with a similar age posterior of $\sim3-9$ Gyr old). Furthermore, DGSAT I shows evidence of having an extended SFH, which might be related to its location in the field.
\item Independently of SED fitting, the optical--NIR colors of VCC 1287 and DGSAT I are significantly different from our comparison dE VCC 1122. VCC 1287 is more metal-poor and DGSAT I is younger and more metal-rich than the dE. 
\item With VCC 1287 and the Coma UDGs \citep{kadowaki17,gu17}, a general picture is emerging where cluster UDGs might be ``failed" galaxies, but the field UDG DGSAT I seems to be more consistent with a feedback-induced expansion scenario \citep{dicintio17}. 
\end{enumerate}

Our work in this paper has focused on only a few objects using a detailed and rigorous Bayesian method. It will be important to expand the sample size of objects analyzed, especially for the purpose of doing a ``differential comparison" between the stellar populations of UDGs and dwarfs. Our Bayesian SED fitting method is complementary to the full spectral fitting method of \citet{gu17} and likely can reach lower limiting surface brightnesses and be more easily applied to a large and diverse sample of UDGs. There are at least three ways to naturally follow up on our work in the future: (1) apply our SED fitting methodology to a larger sample of red UDGs, (2) extend this study to encompass the so-called ``blue UDGs," which may have emission lines and archival UV/IR detections enabling more robust measurements of recent SFHs, ages, metallicities and dust attenuation, and (3) for a statistical sample of UDGs with robust multi-band imaging but unknown distances, derive photometric redshift posteriors while marginalizing over stellar population properties (akin to studies of high-redshift galaxies).

\section*{Acknowledgements}
We greatly thank Laura Ferrarese for providing us with Elixir-LSB CFHT imaging of VCC 1287. We thank James Ingalls for his help with ``de-striping" our IRAC images, and Sean Carey and Tom Jarrett for advice on systematic uncertainties in IRAC imaging. We thank Fangzhou Jiang, Avishai Dekel, Andi Burkert and Soeren Larsen for useful discussions, Johnny Greco and Chris Conselice for helpful comments on the paper, and the anonymous referee for invaluable suggestions. VP acknowledges useful conversations with Arianna di Cintio, Meng Gu, Timothy Carleton, and Tsang Keung Chan at the 2017 Santa Cruz Galaxy Formation Workshop, and separately Rachel Somerville, Joel Primack, Sandy Faber and David Koo. AJR was supported by National Science Foundation grant AST-1616710 and as a Research Corporation for Science Advancement Cottrell Scholar. JPB was supported by NSF AST-1616598. This research was supported in part by the National Science Foundation under Grant NSF PHY11-25915. This work is based in part on observations made with and archival data obtained with the $Spitzer$ Space Telescope, which is operated by the Jet Propulsion Laboratory, California Institute of Technology under a contract with NASA. This research made use of Astropy (particularly \texttt{photutils}), a community-developed core Python package for Astronomy \citep{astropy13}.

\bibliographystyle{apj}
\bibliography{references}

\appendix 

\section{Parameter Covariances with Dust}\label{sec:covdust}
\begin{figure*} 
\begin{center}
\includegraphics[width=\hsize]{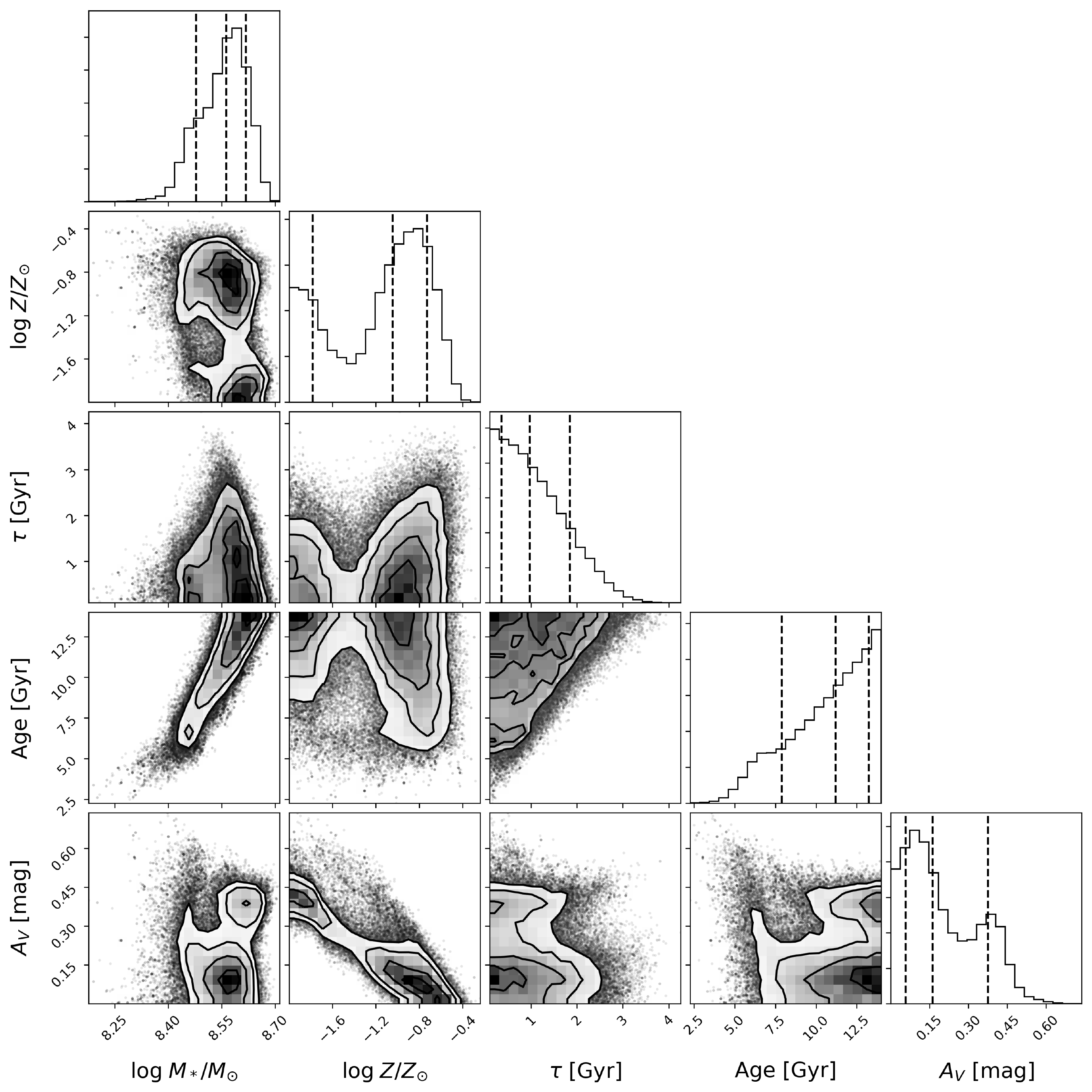}
\end{center}
\caption{Corner plot for the Virgo dE VCC 1122 showing the full covariances between the five free parameters mapped out via MCMC. IRAC4 was excluded from this fit. The vertical dashed lines mark the 16, 50 and 84 percentiles of the posterior distributions.}
\label{fig:cornerVCC 1122}
\end{figure*}

\begin{figure*} 
\begin{center}
\includegraphics[width=\hsize]{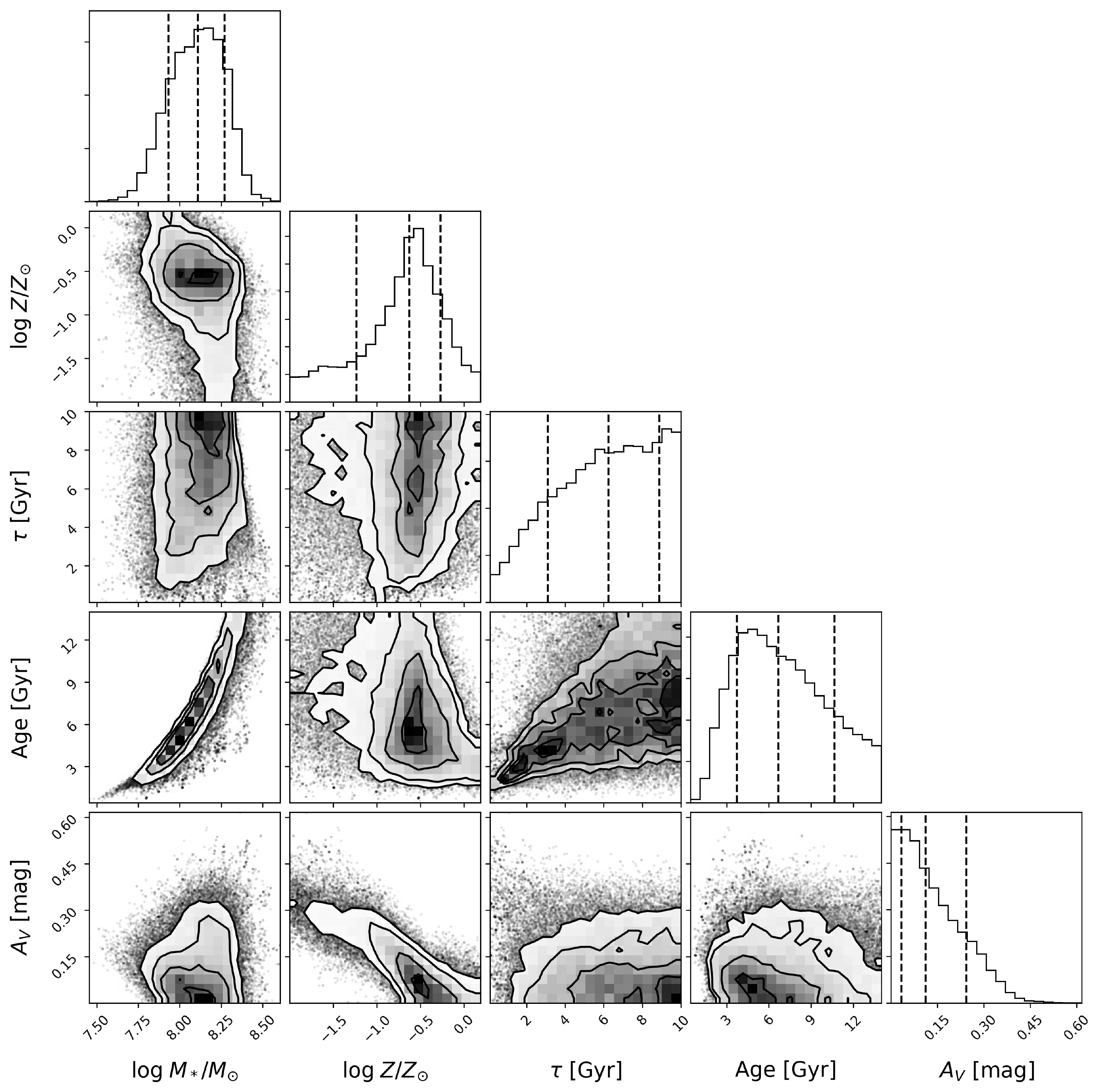}
\end{center}
\caption{Corner plot for the UDG DGSAT I showing the full MCMC-based covariances between the five free parameters. The vertical dashed lines mark the 16, 50 and 84 percentiles of the posterior distributions.}
\label{fig:cornerDGSAT I}
\end{figure*}

\begin{figure*} 
\begin{center}
\includegraphics[width=\hsize]{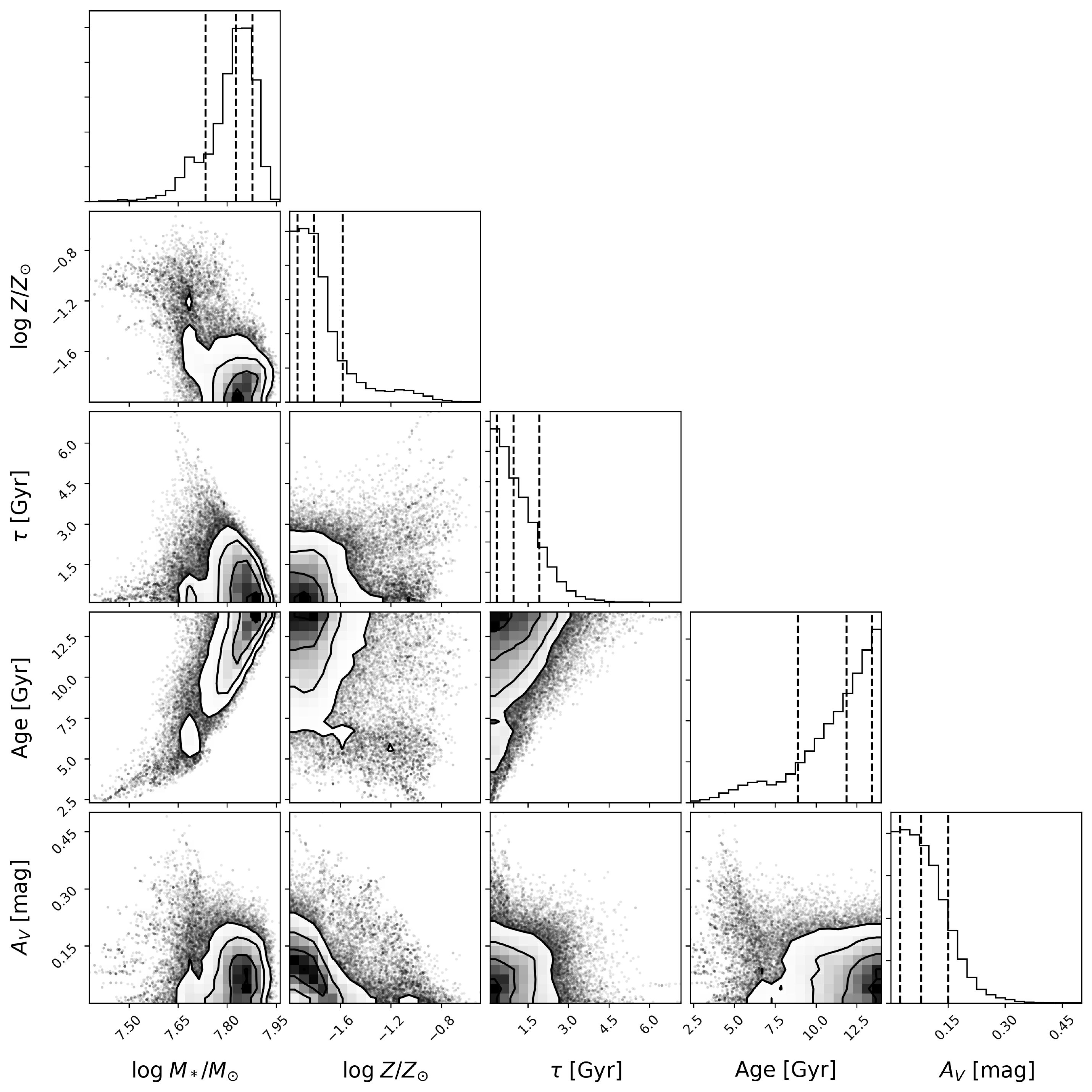}
\end{center}
\caption{Corner plot for the UDG VCC 1287 showing the full MCMC-based covariances between the five free parameters. The vertical dashed lines mark the 16, 50 and 84 percentiles of the posterior distributions.}
\label{fig:corner1287}
\end{figure*}

In \autoref{fig:cornerVCC 1122}, \autoref{fig:cornerDGSAT I}, and \autoref{fig:corner1287}, we show the full covariances between the five free parameters in \texttt{prospector} for VCC 1122, DGSAT I, and VCC 1287, respectively. For these corner plots, we assumed a uniform prior over $A_V=0-4$ mag, and IRAC4 was excluded from the fit for the dE VCC 1122. These covariance plots correspond to the black distributions shown in \autoref{fig:posteriors}.

\section{Parameter Covariances with Minimal/No Dust}\label{sec:covnodust}
\begin{figure*} 
\begin{center}
\includegraphics[width=\hsize]{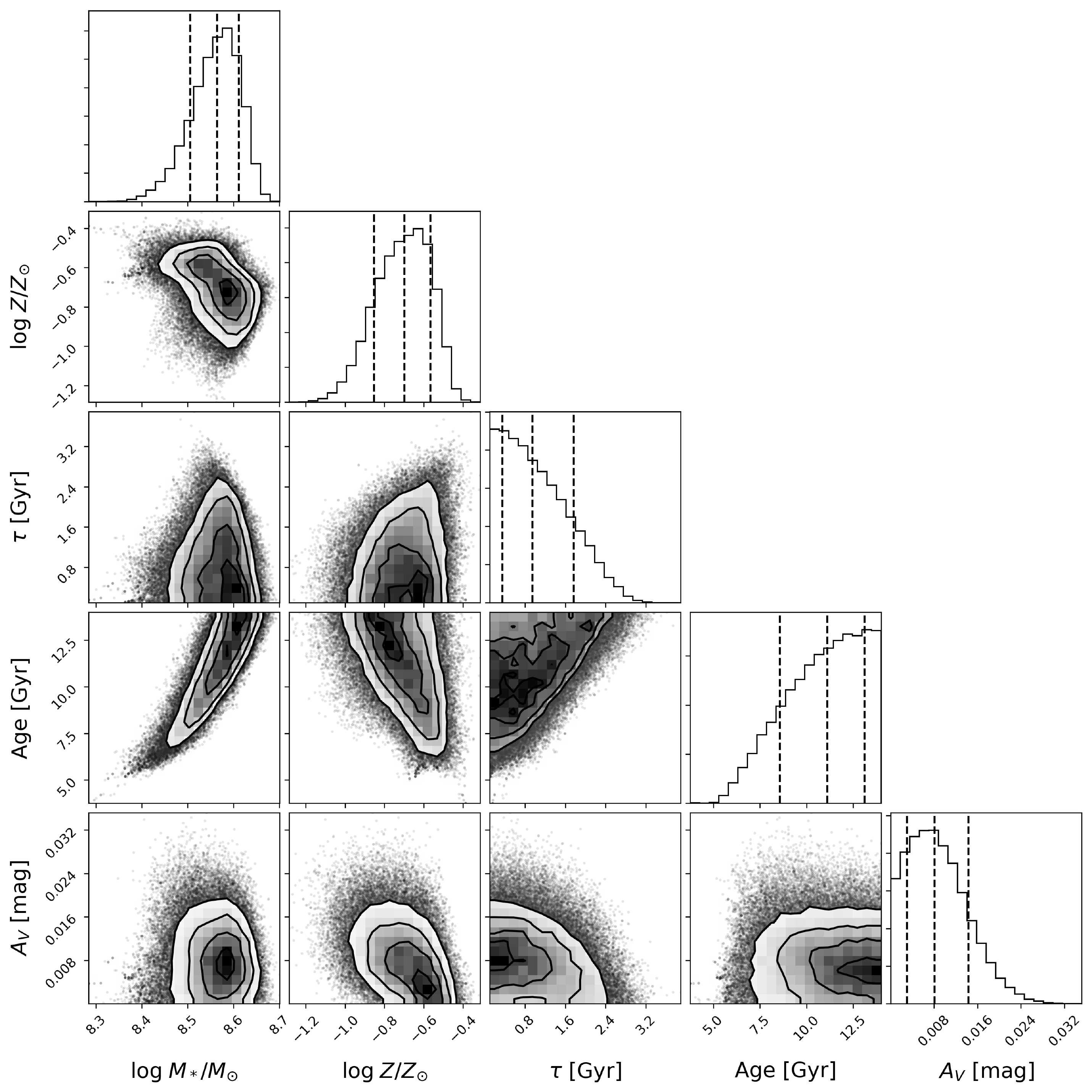}
\end{center}
\caption{Corner plot for the Virgo dE VCC 1122 showing the full covariances between the five free parameters mapped out via MCMC. Here, IRAC4 was included in the fit and helped to rule out high diffuse interstellar dust models compared to \autoref{fig:cornerVCC 1122}. The vertical dashed lines mark the 16, 50 and 84 percentiles of the posterior distributions.}
\label{fig:cornerVCC 1122b}
\end{figure*}

\begin{figure*} 
\begin{center}
\includegraphics[width=\hsize]{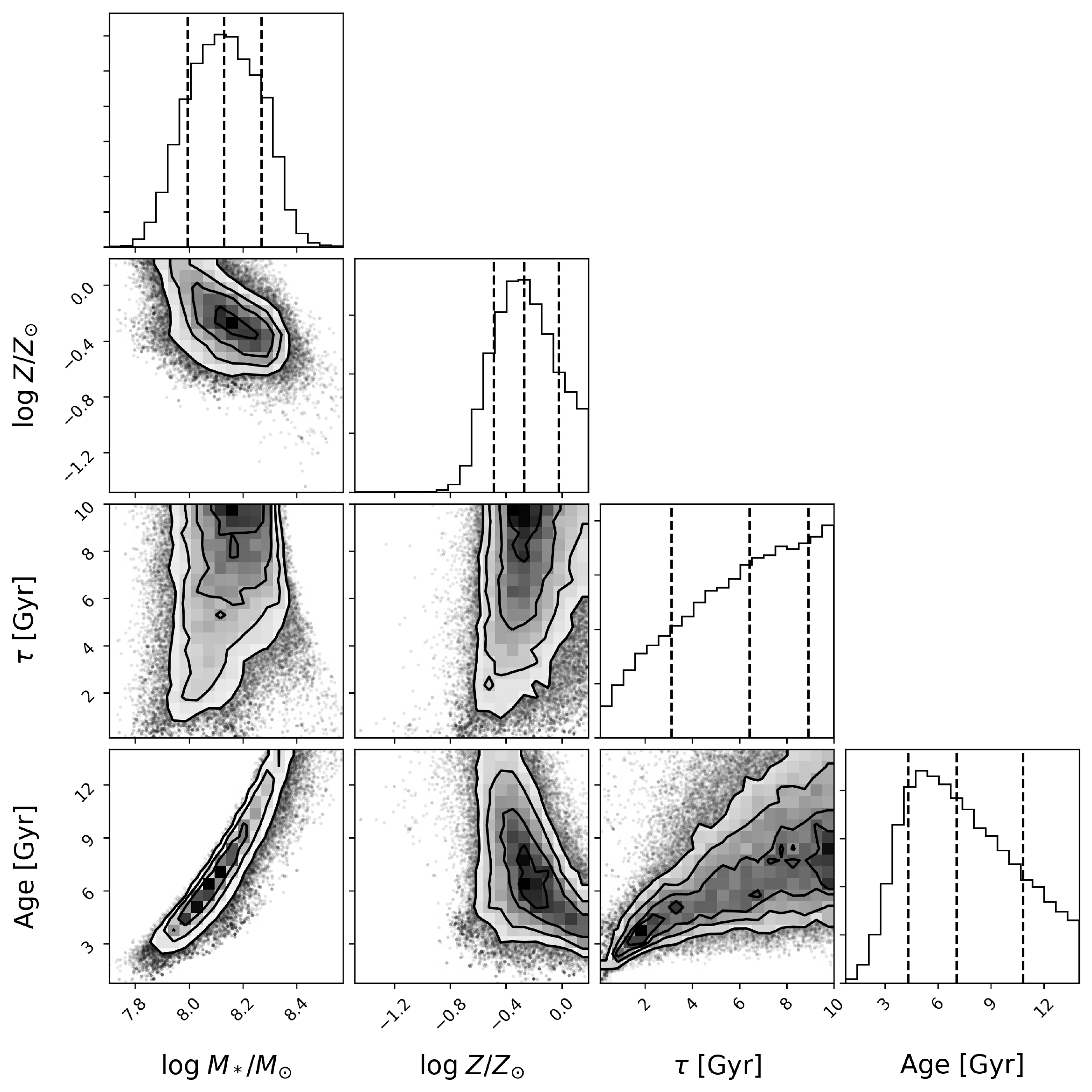}
\end{center}
\caption{Corner plot for the UDG DGSAT I showing the full MCMC-based covariances between the four free parameters, with $A_V$ fixed to $0$ mag (i.e., no diffuse interstellar dust). The vertical dashed lines mark the 16, 50 and 84 percentiles of the posterior distributions.}
\label{fig:cornerDGSAT Ib}
\end{figure*}

\begin{figure*} 
\begin{center}
\includegraphics[width=\hsize]{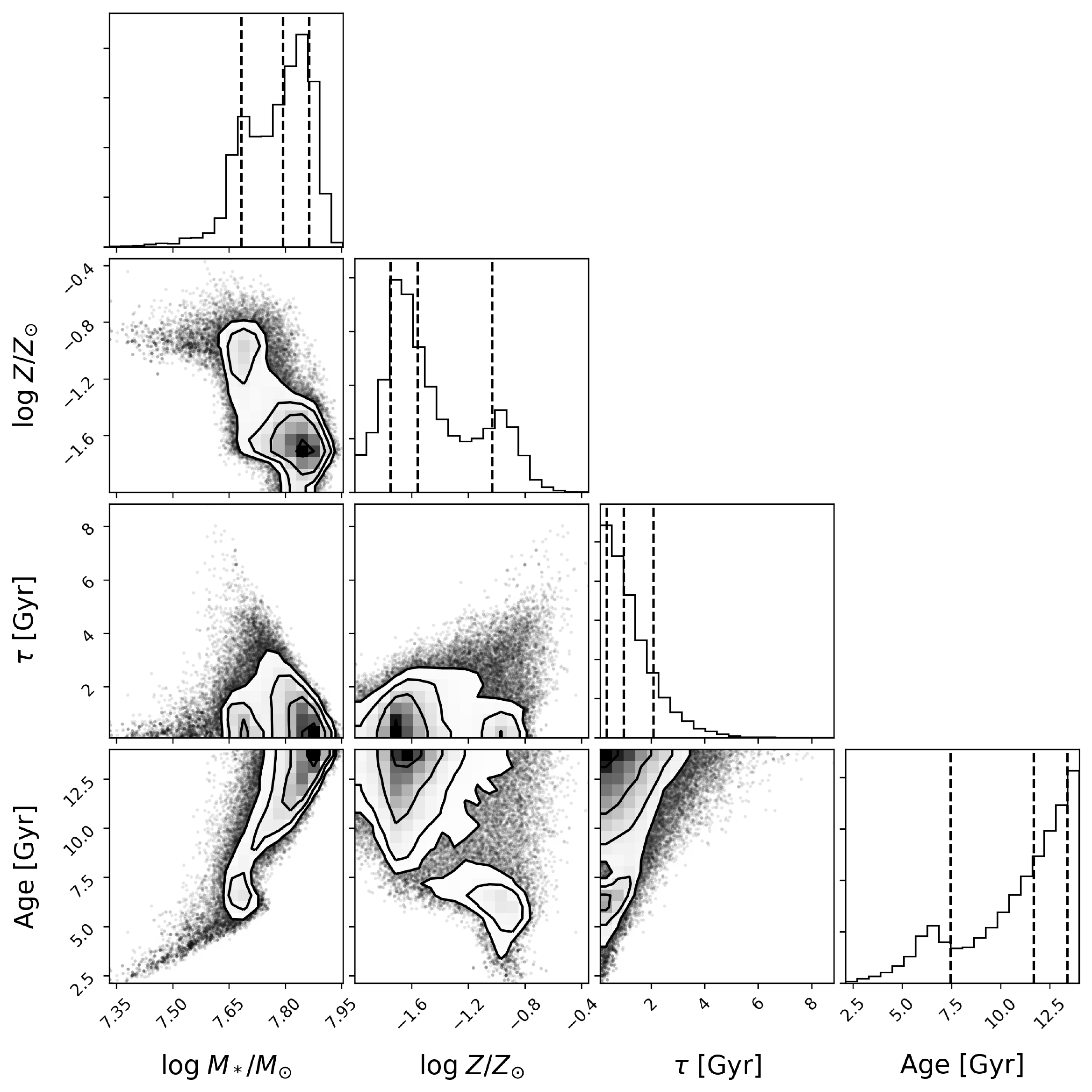}
\end{center}
\caption{Corner plot for the UDG VCC 1287 showing the full MCMC-based covariances between the four free parameters, with $A_V$ fixed to $0$ mag (i.e., no diffuse interstellar dust). The vertical dashed lines mark the 16, 50 and 84 percentiles of the posterior distributions.}
\label{fig:corner1287b}
\end{figure*}

Unlike in Appendix \ref{sec:covdust}, here we include IRAC4 in the fit for the dE VCC 1122, which helps to rule out high diffuse interstellar dust models (continuing to leave $A_V$ as a free parameter, with uniform prior over $0-4$ mag). Since we did not find deep enough archival IRAC4 or MIR/FIR data for the two UDGs, we instead fixed $A_V$ to $0$ mag and fit only for the stellar mass, stellar metallicity, e-folding time and age. The parameter covariances for VCC 1122, DGSAT I, and VCC 1287 are shown in \autoref{fig:cornerVCC 1122b}, \autoref{fig:cornerDGSAT Ib}, and \autoref{fig:corner1287b}, respectively. These covariance plots correspond to the magenta/orange distributions shown in \autoref{fig:posteriors}.

\section{VCC 1287: Comparison of CFHT Public and CFHT Elixir-LSB Imaging}\label{sec:elixir}
\begin{figure*} 
\begin{center}
\includegraphics[width=\hsize]{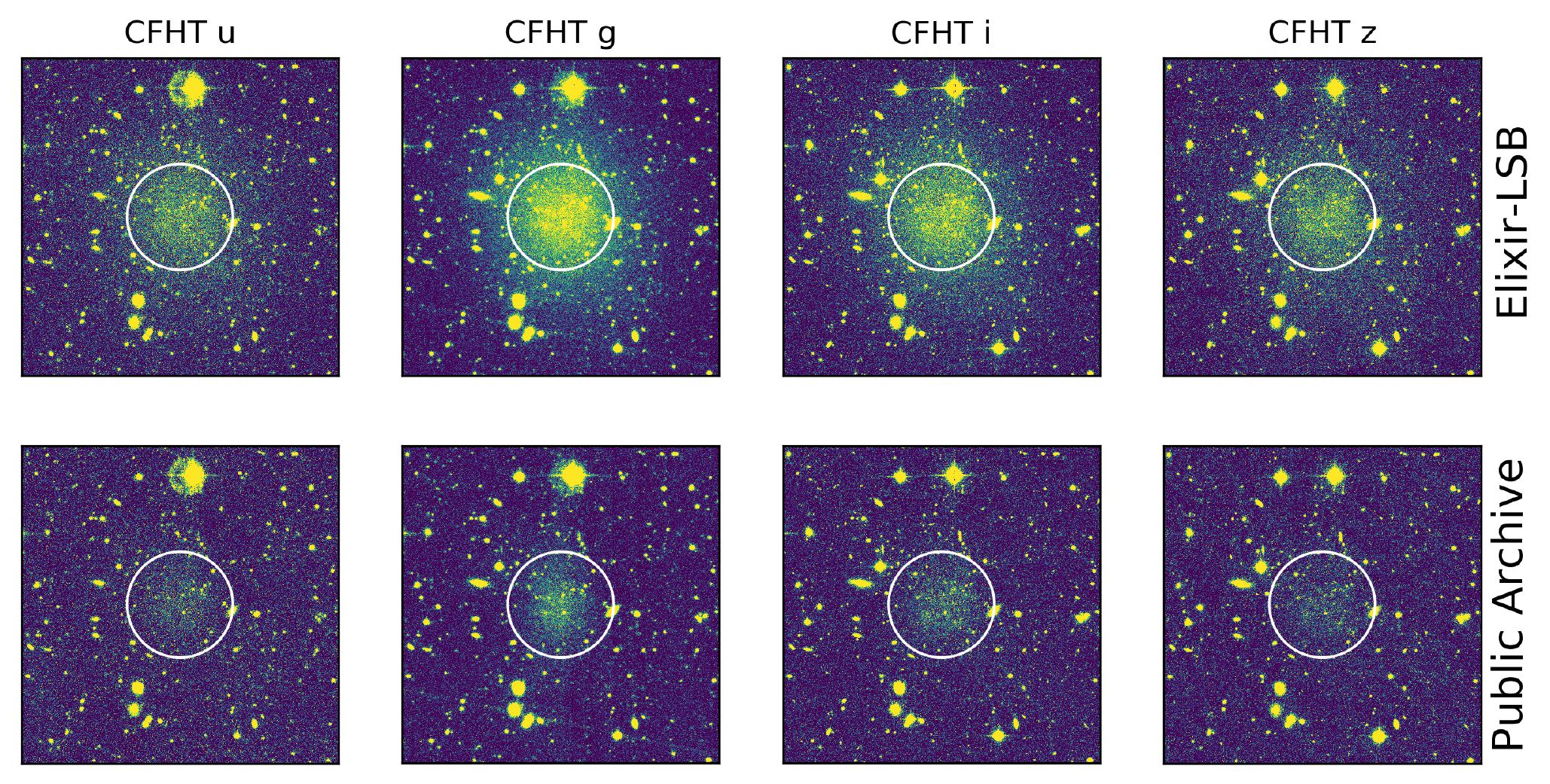}
\end{center}
\caption{A comparison of how VCC 1287 looks in the Elixir-LSB data that we are using (top row) versus in the public archival data (bottom row), for the CFHT $u^*giz$ filters. A significant fraction of UDG light was erroneously subtracted off as part of the background in the public archival imaging, which makes the CFHT public archival data unsuitable for both structural measurements (e.g., half-light radius) and even SED fitting. The white circles show a 30 arcsec aperture within which the UDG is a factor of $\sim3-4$ times brighter in the Elixir-LSB imaging. The colorbar limits have been fixed to span the $50-97$ percentiles of the image array in each subplot such that the galaxy is highlighted relative to the background residuals.}
\label{fig:elixir}
\end{figure*}

In \autoref{fig:elixir}, we show that VCC 1287 looks significantly fainter in the public archival CFHT imaging versus the Elixir-LSB imaging. In the public archival CFHT imaging, the background mesh size is smaller than the size of the UDG and thus a significant fraction of the UDG light is fitted as part of the background and subtracted off \citep{gwyn08}. In contrast, the Elixir-LSB pipeline \citep[see][]{ferrarese12,duc15} is optimized specifically for characterizing and subtracting the background around extended low surface brightness objects like VCC 1287. When we run GALFIT on the Elixir-LSB CFHT imaging, we find S{\'e}rsic half-light radii closer to $\sim40-50$ arcsec \citep[rather than the 30" reported by][]{beasley16}, which is more consistent with the half-light radius of the UDG in IRAC1. 

When measuring the magnitude within the same 30" circular aperture as defined in \autoref{sec:photometry}, there is a large systematic difference between the public archival data and the Elixir-LSB data. Specifically, $\Delta u\approx1.2$ mag, $\Delta g\approx1.3$ mag, $\Delta i\approx1.3$ mag, and $\Delta z\approx1.5$ mag, such that the UDG is artificially $\sim3-4\times$ fainter in the public archival imaging. Since the effect is not constant with bandpass, the SED shape of VCC 1287 itself gets distorted in an unphysical way. Thus the galaxy colors reported in \citet{beasley16} are not suitable for SED fitting. This exercise highlights the need for specialized background subtraction when doing photometry and SED fitting of low surface brightness galaxies, and provides a warning that imaging reduced with standard pipelines may not be suitable for these purposes.

\citet{beasley16} reported total $g=17.8$ AB mag which would correspond to a total stellar mass of $\sim2\times10^7M_{\odot}$ assuming our own \texttt{prospector}-based stellar $M_*/L_g=1.6$ (this is measured within a 30 arcsec aperture, which is their $R_e$). Our aperture stellar mass is itself already $\sim3\times$ higher than that, and our total stellar mass is a factor of ten even higher.

\section{GALFIT Results}\label{sec:galfit}
\begin{figure*} 
\begin{center}
\includegraphics[width=\hsize]{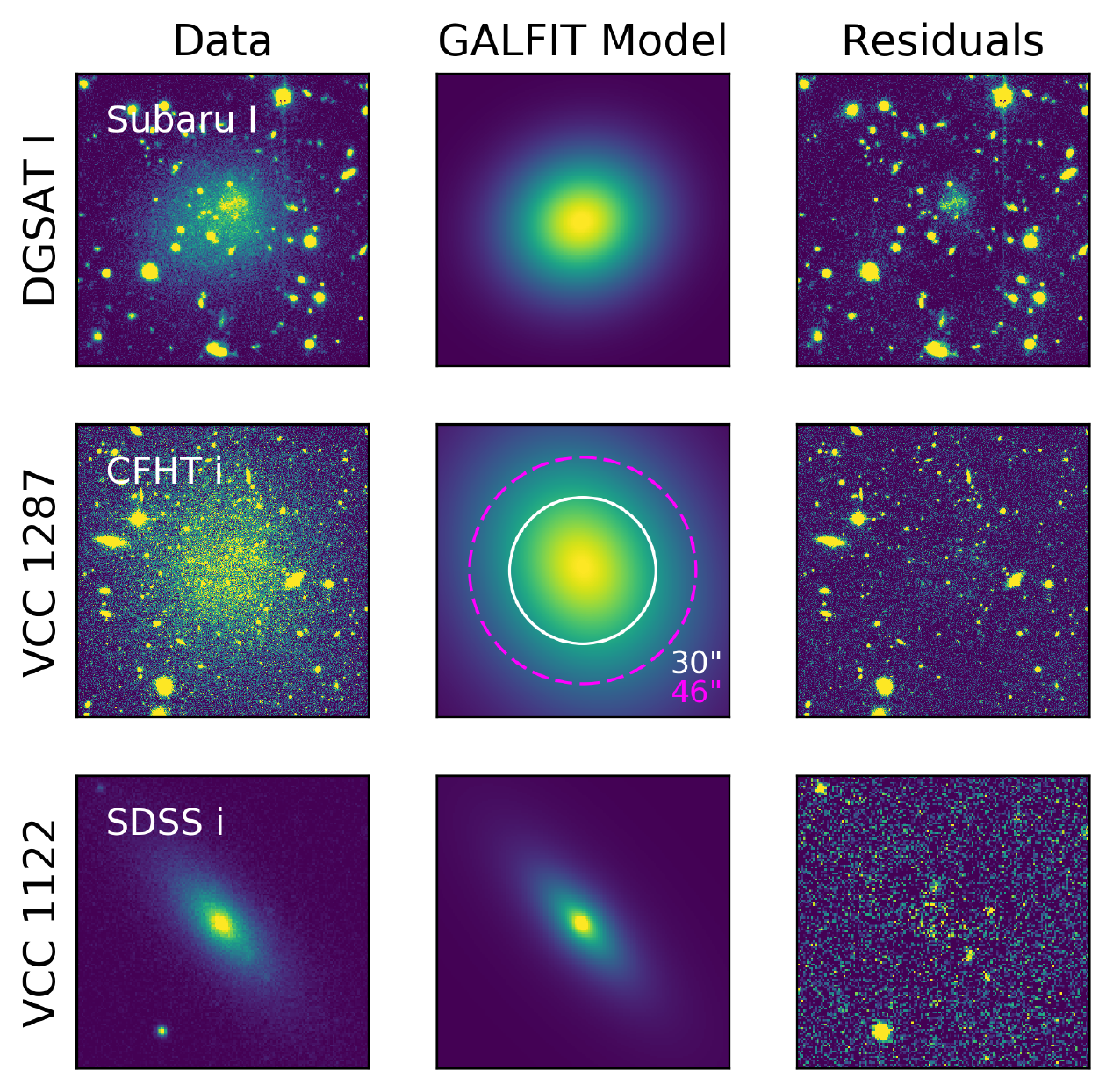}
\end{center}
\caption{Optical imaging, GALFIT model, and residuals for our three galaxies. The bandpasses are $I$ for DGSAT I, CFHT-$i$ for VCC 1287, and SDSS-$i$ for VCC 1122. For VCC 1287, we also show in white a 30 arcsec circle corresponding to the half-light radius found by \citet{beasley16} using the available archival CFHT imaging that was not optimized for low surface brightness features, and a 46.4 arcsec circle in magenta showing our revised optical half-light radius measurement.}
\label{fig:galfit}
\end{figure*}

In \autoref{fig:galfit}, we show the optical images that we ran GALFIT on, the GALFIT models themselves, and the residuals for all three of our galaxies. The bandpasses chosen were $I$ for DGSAT I, CFHT-$i$ for VCC 1287, and SDSS-$i$ for VCC 1122. All of our GALFIT models and residuals are reasonable. For DGSAT I and VCC 1287, a single S{\'e}rsic profile was adequate to capture the smooth galaxy light. We let all parameters vary: centroid, total enclosed magnitude, effective radius, S{\'e}rsic index $n$, axis ratio $b/a$, and position angle. 

For DGSAT I, our best-fit parameters are in good agreement with those of \citet{martinezdelgado16}; both they and we masked out the ``overdensity" when fitting. Specifically, we found total $I=17.70$ AB mag (after applying the MW reddening correction) and $R_e=13.4$ arcsec. \citet{martinezdelgado16} found $R_e=12.5$ arcsec and total $I=17.17$ Vega mag; applying the AB to Vega transformation from Table 1 of \citet{blanton07}, we get $I=17.25$ mag, which agrees within 0.1 mag. Any minor differences between our GALFIT results and those of \citet{martinezdelgado16} can likely be ascribed to our different masking algorithms (theirs was done iteratively with additional masking ``by hand"). The formal errors on our GALFIT free parameters are negligible.

For VCC 1287, our best-fit S{\'e}rsic parameters are significantly different from those reported by \citet{beasley16} because the UDG was over-subtracted in their CFHT imaging (see Appendix \ref{sec:elixir}). We find total $i=15.05$ AB mag (after the MW reddening correction) and $R_e=46.4$ arcsec. \citet{beasley16} measured $R_e=30.2$ arcsec, which is nearly a factor of two smaller. Furthermore, using the $M_g$ and $(g-i)_0$ values they give in their section 2.1, we calculate that their total $i=16.85$ AB mag (after applying the MW reddening correction), which is nearly two magnitudes fainter than our measurement. The formal errors on our GALFIT free parameters are negligible.

For the dE VCC 1122, we required three structural components in GALFIT: two S{\'e}rsic profiles and one exponential disk. We left all S{\'e}rsic parameters free for both profiles (as above), and we also left all the exponential disk parameters free: centroid, total enclosed magnitude, exponential scale radius, axis ratio, and position angle. If we only allowed a single S{\'e}rsic profile, or one S{\'e}rsic plus the exponential disk, significant residuals were left over (particularly in the ``wings" of the dE). It is common for dEs to require multiple structural components \citep[e.g.,][]{janz14}. For the first S{\'e}rsic component, we find $i=14.74$ AB mag (no MW reddening correction yet) and $R_e=5.0$ arcsec. For the second S{\'e}rsic component, we find $i=16.63$ AB mag and $R_e=1.7$ arcsec. For the exponential disk component, we find $i=14.21$ AB mag and $r_s=8.4$ arcsec. Adding up the fluxes and correcting for MW reddening, we find that the total $i=13.59$ AB mag. The exponential disk component is the most extended and luminous component (and helps to account for the ``wings"). The formal errors on all free parameters are negligible, with the exception of the PA for the second S{\'e}rsic component ($15.12\pm19.51$ deg, but this is relatively unimportant). 

\end{document}